\documentclass[a4paper,11pt]{article}

\pdfoutput=1

\usepackage{jheppub}
\usepackage{mathrsfs}
\usepackage{xspace}
\usepackage{times}
\usepackage{etoolbox}
\usepackage{animate}

\renewcommand{\vec}[1]{\mathbf{#1}}
\newcommand{\C}{\textit{C}\xspace}
\newcommand{\CP}{\textit{CP}\xspace}
\newcommand{\MS}{$\overline{\rm MS}$\xspace}
\renewcommand{\Re}{\mbox{Re}}
\renewcommand{\Im}{\mbox{Im}}
\newcommand{\Tr}{{\rm Tr}}
\newcommand{\tr}{{\rm tr}}
\newcommand{\Meff}{\mathbf{M}}
\newcommand{\Geff}{\mathbf{\Gamma}}
\newcommand{\sign}{{\rm sign}}
\newcommand{\eqn}{eq.\,}

\makeatletter
\patchcmd{\maketitle}{\@fpheader}{}{}{}
\makeatother

\preprint{MPP-2014-40, NSF-KITP-14-087}

\title{Leptogenesis in \emph{crossing} and \emph{runaway} regimes }

\author{A. Hohenegger$^{a}$ and}
\emailAdd{andreas.hohenegger@uis.no}

\author{A. Kartavtsev$^{b}$}
\emailAdd{alexander.kartavtsev@mpp.mpg.de}

\affiliation[a]{University of Stavanger, Kjell Arholms gate 41, 4036 Stavanger, Norway}
\affiliation[b]{Max-Planck-Institut f\"ur Physik, F\"ohringer Ring 6, 80805 M\"unchen, Germany}

\arxivnumber{1404.5309}

\abstract{We study the impact of effective thermal masses and widths on resonant leptogenesis. 
We identify two distinct possibilities which we refer to as `crossing' and `runaway' regimes.   
In the runaway regime the mass difference grows monotonously with temperature, whereas it initially 
decreases in the crossing regime, such that the effective masses become equal at some temperature. 
Following the conventional logic the source of the asymmetry would vanish in the latter case. 
Using non-equilibrium quantum field theory, we analytically demonstrate that the vanishing of 
the difference of the effective masses  does however neither imply a suppression nor a strong 
enhancement of the source for the lepton asymmetry. In the vicinity of the crossing point the  
asymmetry calculated in an (improved) Boltzmann limit develops a spurious peak, which signals 
the breakdown of the quasiparticle approximation. In the exact result this spurious enhancement 
is compensated by coherent transitions between the two mass shells. Despite the breakdown of the 
quasiparticle approximation off-shell contributions remain negligibly small even at the crossing point.}

\begin{document}

\maketitle
\flushbottom
\newpage

\section{\label{Introduction}Introduction}

From the theoretical point of view the baryogenesis via leptogenesis scenario 
\cite{Fukugita:1986hr} is a very attractive explanation for the observed baryon 
asymmetry of the Universe. One of its key ingredients are heavy Majorana neutrinos, 
which may cause a lepton asymmetry to emerge. 
The generation of the asymmetry can proceed via \CP-violating decays and inverse 
decays of the heavy neutrinos \cite{Fukugita:1986hr}, their \CP-violating oscillations 
\cite{Akhmedov:1998qx}, or via a combination of the two. The 
first case is typically realized for Majorana neutrinos 
with masses considerably larger than the sphaleron freeze-out temperature. This possibility has been 
explored extensively using the usual Boltzmann-like equations with decay and scattering 
amplitudes computed using methods of zero temperature \cite{Luty:1992un,Flanz:1994yx,
Covi:1996wh,Roulet:1997xa,Davidson:2002qv,Buchmuller:2002rq,
Buchmuller:2003gz,DiBari:2004en,Abada:2006ea,Blanchet:2007hv,JosseMichaux:2007zj,
Blanchet:2008pw} or thermal \cite{Covi:1997dr,Giudice:2003jh,Kiessig:2009cm,Garny:2010nj} 
quantum field theory. The second case is typically realized for  
Majorana neutrinos with masses below the sphaleron freeze-out temperature. It 
has been studied  using the 
`density matrix formalism' \cite{Shaposhnikov:2008pf,Boyarsky:2009ix,Shaposhnikov:2009zzc,
Canetti:2010aw,Asaka:2011wq,Canetti:2012vf,Canetti:2012zc,Shuve:2014zua,Dev:2014laa} which 
was originally developed in \cite{Sigl:1992fn} and cross-checked in an alternative approach 
\cite{Gagnon:2010kt}. 

Recently, various aspects of leptogenesis have been re-analysed using the first-principle 
Kada\-noff--Baym formalism \cite{Buchmuller:2000nd,Anisimov:2008dz,Anisimov:2010dk,
Anisimov:2010aq} as well as   self-consistent Boltzmann-like \cite{Garny:2009rv,
Garny:2009qn,Beneke:2010wd,Garbrecht:2010sz,Frossard:2012pc,Frossard:2013bra} and 
qu\-a\-n\-tum-kinetic  \cite{DeSimone:2007rw,Beneke:2010dz,Drewes:2012ma,Garbrecht:2011aw}
equations systematically derived from the former. A lot of effort has been put into the 
analysis of the phenomenologically particularly interesting 
scenario of resonant leptogenesis \cite{Flanz:1996fb,Covi:1996fm,Pilaftsis:1997jf,
Pilaftsis:1997dr,Pilaftsis:1998pd,Buchmuller:2000as,Pilaftsis:2003gt,Pilaftsis:2005rv,
Anisimov:2005hr}.  
Resonant leptogenesis is realized for a quasi-degenerate mass spectrum of the heavy 
neutrinos, when the difference of the  masses is comparable to the sum of the
decay widths. In this domain of the parameter space the \CP-violating parameters 
are resonantly enhanced. 
In the case of leptogenesis via \CP-violating decays and inverse decays this allows 
one to lower the Majorana masses down to the TeV scale 
\cite{Pilaftsis:2003gt,Pilaftsis:2005rv} which is in principle accessible at the 
LHC \cite{Blanchet:2009bu,Ibarra:2011xn,Dev:2013wba,Deppisch:2013jxa,Canetti:2014dka,
Das:2014jxa}. In case 
of leptogenesis via \CP-violating oscillations of the Majorana neutrinos their 
masses can be as low as a few GeV \cite{Shaposhnikov:2008pf,Boyarsky:2009ix,
Shaposhnikov:2009zzc,Canetti:2010aw,Asaka:2011wq,Canetti:2012vf,Canetti:2012zc,
Shuve:2014zua}, such that they can be searched for in dedicated high-intensity 
experiments \cite{Bonivento:2013jag}.

The regime of resonant asymmetry generation is not only of particularly high interest 
but presumably also the most difficult one to study. Early works using the Boltzmann 
equation  and relying on the zero temperature  quantum field theory concentrated on 
the derivation of \CP-violating parameters $\epsilon_i$. In case of two generations 
of Majorana neutrinos it has been found in \cite{Anisimov:2005hr}, using the on-shell 
renormalization scheme, that in the basis where the mass matrix is diagonal: 
\begin{align}
\label{EpsilonVacuum}
\epsilon_i\propto  \frac{M_j^2-M_i^2}{\bigl(M_j^2-M_i^2-\frac1{\pi}\ln(M_j^2/M_i^2)\bigr)^2 +\bigl(M_j\Gamma_j-M_i\Gamma_i\bigr)^2}\,.
\end{align}
In the limit of vanishing mass difference, $M_2=M_1$, the numerator of \eqref{EpsilonVacuum}
becomes zero whereas the denominator does not if $\Gamma_2\neq \Gamma_1$. The vanishing of the 
\CP-violating parameters is required in this limit, because the corresponding Lagrangian is 
\CP-invariant \cite{Kniehl:1996bd,Hohenegger:2013zia}.
On the other hand, if $M_2\rightarrow M_1$ and $\Gamma_2\rightarrow \Gamma_1$ simultaneously 
then, according to the above expression, the \CP-violating parameters do not vanish even 
though the Lagrangian is  \CP-invariant as well in this case. The origin of this problem 
lies in the use of the quasiparticle picture built-in in the Boltzmann approximation. 
The very fact that the peaks 
of the spectral functions that correspond to the quasiparticle excitations strongly overlap 
in the resonant regime renders the use of Boltzmann equations problematic. The question which 
behaviour the \CP-violating parameter exhibits in this `doubly degenerate' limit has recently 
been answered in \cite{Garny:2011hg}, using the formalism of non-equilibrium quantum field 
theory and without invoking the quasiparticle approximation. It 
has been found  that, for $M_2\rightarrow M_1$ and $
\Gamma_2\rightarrow \Gamma_1$, the regulator $M_j\Gamma_j-M_i\Gamma_i$ in the denominator of 
\eqref{EpsilonVacuum} is effectively replaced by $M_j\Gamma_j
+M_i\Gamma_i$ due to additional contributions that describe coherent transitions between 
the Majorana neutrino species. This implies that in the resonant regime both \CP-violating
(inverse) decays \emph{and} oscillations play an important role and must be taken into 
account in a self-consistent analysis.  

\begin{figure}
   \includegraphics[width=0.49\textwidth]{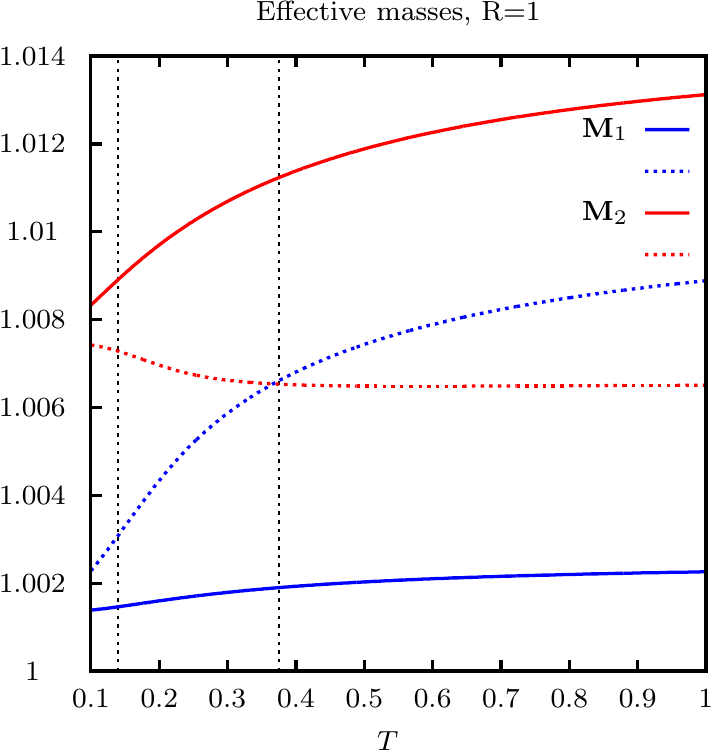}
   \,
   \includegraphics[width=0.49\textwidth]{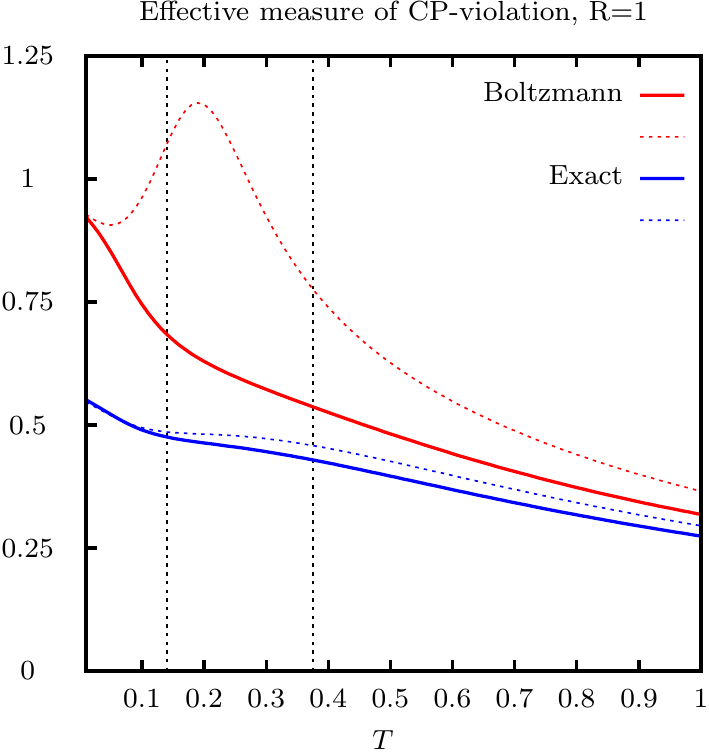}
\caption{\label{Illustration} Sketch of the dependence of the effective masses (left) and 
effective measure of \CP-violation (right) in the runaway (solid lines) and crossing (dotted 
lines) regimes  for a quasi-degenerate mass spectrum of the heavy particles. $R$ denotes 
the resonance parameter which quantifies the mass splitting in vacuum (see below).}
\end{figure}

According to the conventional analysis the \CP-violating parameters take their maxima if the 
mass difference is of the order of the sum of the decay widths. However the
 early Universe expands and cools rapidly. During the time interval 
in which most of the asymmetry is generated, the temperature (measured in units of the heavy
neutrino mass) can drop substantially. In the favoured regime, thermal corrections to the 
effective masses can be  comparable to the mass difference itself. Depending on the values 
of the couplings there are two possibilities, see figure \ref{Illustration} (left). 
In the \emph{runaway} regime the mass difference grows with increasing temperature, 
whereas in the \emph{crossing} regime the difference of the masses initially decreases
such that the two masses become equal at some temperature, and then increases again 
with increasing temperature. In this work we study the influence of 
thermal corrections to the masses in these two regimes, which have not been considered before,
on resonant leptogenesis.
To this end we consider a simple toy-model which proved to be useful in the past for the 
analysis of leptogenesis  \cite{Garny:2009rv,Garny:2009qn,Garny:2010nj,Garny:2010nz,Hohenegger:2013zia}. 
Throughout this work we emphasize the strict requirement that the obtained source terms for the lepton 
asymmetry need to respect the \CP-properties of the underlying theory represented by the Lagrangian 
as the parameters of corresponding 
quasiparticles (insofar they can be defined) evolve due to medium effects. 

In order to be able to study the interplay of coherent oscillations and resonant enhancement 
rigorously we also adopt a simplified physical picture \cite{Anisimov:2008dz,Anisimov:2010dk,Anisimov:2010aq,Garny:2011hg}. 
We neglect the expansion of the Universe and assume instead that the toy-Majorana neutrinos are 
coupled through their decays to a thermal bath composed of the decay products (toy-leptons). 
The deviation from thermal equilibrium needed to produce an asymmetry 
 is induced by an instantaneous perturbation. This setting may differ from the conventional 
 physical picture connected to standard cosmology but allows us to obtain analytic solutions 
 and study the source of the asymmetry generation rigorously from first principles, in terms 
 of statistical propagators and spectral functions.

Before we present the qualitatively interesting results obtained 
in this approach let us discuss 
what one would \emph{naively} expect from the vacuum expression \eqref{EpsilonVacuum}
in both regimes. Because in the runaway regime the difference of the effective masses 
grows  with temperature, one could expect that the overlap of the peaks of the spectral 
function decreases simultaneously. Therefore, the quality of the quasiparticle 
approximation can be expected to improve. The growing mass difference 
is also expected to result in a smaller \CP-violation. On the other hand, because in the 
crossing regime the difference of the effective masses vanishes at some temperature, the 
peaks of the spectral function are expected to overlap at this point. This suggests  
a complete breakdown of the quasiparticle approximation and large relative size of off-shell
contributions. Furthermore, from \eqref{EpsilonVacuum} 
one could also expect that the \CP-violating parameters vanish, or are at least suppressed at 
the crossing point. At even higher temperatures the mass difference grows again and 
the applicability of the quasiparticle approximation may be expected to be restored. 
The same logic would imply that in the $M_j = M_i$ case $\epsilon_i$ 
vanishes at zero temperature but would in general evolve into a finite one as the 
temperature is increased.

The results of the first-principles analysis, see figure \ref{Illustration} (right), show that 
this naive picture is only partially correct. In the runaway regime we observe 
that the result obtained in the Boltzmann approximation slowly approaches the exact one at 
high temperatures. This signals that the quality of the quasiparticle approximation 
indeed improves at high temperatures. In agreement with the expectations the 
effective measure of \CP-violation decreases monotonously with increasing temperature. In the 
crossing regime the quasiparticle approximation breaks down at the crossing point. As 
can be inferred from figure \ref{Illustration} (right), at the crossing temperature the 
\CP-violating source computed in an improved Boltzmann approximation develops a spurious peak  
which is absent in the exact result. 
On the other hand, the expected vanishing (or at least suppression) of the 
source at the crossing point does not take place.

The outline of the paper is as follows. In section \ref{Setup} we present the toy-model and  
derive an equation for the asymmetry in the framework of 
non-equilibrium quantum field theory. In section \ref{FundamentalSymmetries} we 
analytically demonstrate that the asymmetry automatically vanishes if both the Lagrangian 
and initial conditions are \CP-symmetric. The analysis of the effective masses and 
widths as well as of the behaviour of the spectral function is carried out in section 
\ref{TwoRegimes}. In section \ref{BreitWigner} we present analytical estimates of 
the leading contributions to the effective measure of \CP-violation.
Numerical estimates of the size of the sub-leading contributions are given in section
\ref{Numerics}. We summarize the main results in the beginning of each section. Finally, 
in section \ref{Summary} we conclude 
and give a qualitative explanation for the difference between the naively expected 
behaviour and the exact results.

\section{\label{Setup}Setup}

In this section we derive an equation for the asymmetry in the framework of 
non-equilibrium quantum field theory. The derivation closely follows the analysis which was 
performed in \cite{Garny:2011hg} and recently generalized to the case of expanding 
universe in \cite{Iso:2013lba,Iso:2014afa}. We also establish a connection between this 
first-principle approach and the commonly used Boltzmann approximation. In addition 
we demonstrate that in the Boltzmann approximation the test solution that we use 
corresponds to the weak washout regime and describes free decay of the heavy particles.

\paragraph{Model.}

To reduce the technical complications to a minimum and yet to include all qualitatively important effects  
for the generation of the asymmetry we use a simple toy model studied previously in 
\cite{Garny:2009rv,Garny:2009qn,Garny:2010nj,Garny:2010nz,Hohenegger:2013zia}.
The model contains one complex and two real scalar fields:
\begin{align}
   \label{Lagrangian}
   {\cal L} & = \frac12 \partial^\mu\psi_{i}\partial_\mu\psi_{i}
              - \frac12  \psi_{i} M^2_{ij} \psi_{j} 
             + \partial^\mu \bar{b}\partial_\mu b -m^2\, \bar{b} b 
              - \frac{\lambda}{2!2!}(\bar b b)^2
             - \frac{h_{i}}{2!}\psi_{i} bb 
              - \frac{h^*_{i}}{2!}\psi_{i} \bar{b}\bar{b}\,,
\end{align}
where $\bar{b}$ denotes the complex conjugate of $b$.
Here and in the following we assume summation over repeated 
indices, unless otherwise specified.
Despite its simplicity, the model incorporates all features relevant 
for leptogenesis. The real scalar fields imitate the (two lightest) 
heavy right-handed neutrinos, whereas the complex scalar field models 
the leptons. The $U(1)$ symmetry, which we use to define ``lepton'' number, 
is explicitly broken by the presence of the last two terms, just as the 
$B-L$ symmetry is explicitly broken by Majorana mass terms in phenomenological 
models. Thus the first Sakharov condition is fulfilled. 
The couplings $h_i$ model the complex Yukawa couplings of the right-handed 
neutrinos to leptons and the Higgs. By rephasing the complex scalar field at 
least one of the couplings $h_i$ can be made real. If 
${\rm arg}(h_1)\neq {\rm arg}(h_2)$ the other one remains complex and 
there is  \C-violation, as is required by the second Sakharov
condition. Note that, in the scalar toy model, \CP-transformations on 
the fields are identical to \C-transformations up to the sign change of the spatial coordinates.

\paragraph{Non-equilibrium quantum field theory approach.}

As can be inferred from \eqref{Lagrangian} the Noether current of 
the complex field is given by
\begin{align}
j_\mu&=2i\bigl[\,\bar{b}(x)\partial_{\mu_x} b(x)-b(x)\partial_{\mu_x}\bar{b}(x)\bigr]
=2i\lim_{y\rightarrow x}\bigl[\partial_{\mu_x} \bar{b}(y) b(x)-\partial_{\mu_y} b(x)\bar{b}(y)\bigr]\,,
\end{align}
where $b$ and $\bar b$ are field operators in the Heisenberg representation. The expectation 
value of the current with respect to the initial state is
\begin{align}
\label{Current}
J_\mu(x)&=\langle\, j_\mu(x) \rangle  
=2i\lim_{y\rightarrow x}\bigl[\partial_{\mu_x} D_<(x,y)-\partial_{\mu_y} D_>(x,y)\bigr]\,,
\end{align}
where 
\begin{align}
\label{WightmannFunctions}
D_>(x,y)\equiv \langle b(x)\bar{b}(y) \rangle=
		\Tr[\mathscr{P}\,b(x)\bar{b}(y)]\,,\,\,
D_<(x,y)\equiv \langle \bar{b}(y) b(x) \rangle=
		\Tr[\mathscr{P}\,\bar{b}(y)b(x)]\,,
\end{align}
are so-called Wightmann two-point functions. In general, $D_\gtrless$ are complex-valued.
Using the hermiticity of the den\-sity matrix $\mathscr{P}$ and cyclic invariance of the 
trace  we find that they satisfy \cite{Garny:2009rv}
\begin{align}
	\label{gtrless_properties}
	D^*_>(x,y)=D_>(y,x)\,,\quad D^*_<(x,y)=D_<(y,x)\,.
\end{align}
Instead of the Wightmann two-point functions one frequently uses the 
spectral function and statistical propagator:
\begin{align}
\label{WightmannStatSpectr}
D_\gtrless(x,y)&=D_F(x,y)\mp\frac{i}{2}D_\rho(x,y)\,.
\end{align}
As can be inferred from \eqref{WightmannFunctions} and \eqref{WightmannStatSpectr},
they are defined as 
\begin{align}
\label{StatSpectr}
D_F(x,y)\equiv \textstyle{\frac12}\langle \{b(x),\bar{b}(y)\} \rangle\,,\,\,
D_\rho(x,y)&\equiv i \langle [b(x),\bar{b}(y)] \rangle\,,
\end{align}
where the square brackets denote the commutator and the curly ones denote the 
anti-commutator of the fields.  Using \eqref{gtrless_properties} we find that under complex 
conjugation they transform as
\begin{align}
	\label{DFDrhoProp}
	D^*_F(x,y)=D_F(y,x),\quad D^*_\rho(x,y)=-D_\rho(y,x)\,.
\end{align}
Substituting \eqref{WightmannStatSpectr} into \eqref{Current} we obtain
\begin{align}
\label{SimplifiedCurrent}
J_\mu(x)&=2i\lim_{y\rightarrow x}(\partial_{\mu_x}-\partial_{\mu_y})D_F(x,y)
-\lim_{y\rightarrow x}(\partial_{\mu_x}+\partial_{\mu_y})D_\rho(x,y)\,.
\end{align}
The definition of the spectral function, \eqn\eqref{StatSpectr}, combined 
with the canonical equal-time commutation relations, 
\begin{align}
\bigl[b(t,\vec x),\dot{\bar{b}}(t,\vec y)\bigr]=
\bigl[\bar{b}(t,\vec x),\dot{b}(t,\vec y)\bigr]=i\delta(\vec x-\vec y)\,,
\end{align}
then implies that the spectral function does not con\-tri\-bu\-te to the current.
The divergence of the current,
\begin{align}
\label{CurrentDivergence}
\partial^\mu J_\mu(x)&=i\lim_{y\rightarrow x}(\partial^{\mu_x}+\partial^{\mu_y})
(\partial_{\mu_x}-\partial_{\mu_y})D_F(x,y)
=i\lim_{y\rightarrow x}(\square_x-\square_y)D_F(x,y)\,,
\end{align}
can be rewritten using the Kadanoff-Baym equations (KBE) for the complex field. 
For Gaussian  initial conditions the latter take the form \cite{Garny:2009rv}
\begin{subequations}
\label{KBforComplexField}
\begin{align}
	\label{DFequation}
	[\square_x+m^2]D_F(x,y)&=
	\int\limits^{y^0}_{t_0} d^4 z\,\Sigma_F(x,z)D_\rho(z,y)
	-\int\limits^{x^0}_{t_0} d^4 z\,\Sigma_\rho(x,z)D_F(z,y)\,,\\
	\label{Drhoequation}
	[\square_x+m^2]D_\rho(x,y)&=
	\int\limits_{x^0}^{y^0} d^4 z\,\Sigma_\rho(x,z)D_\rho(z,y)\,,
\end{align}
\end{subequations}
where $t_0$  is the initial time surface and  $\Sigma_{F(\rho)}$ are 
the statistical (spectral) components of the self-energy.
Substituting \eqref{DFequation} into \eqref{CurrentDivergence} we obtain
\begin{align}
\label{MasterEquation0}
\partial^\mu J_\mu(x)=-i\int\limits^{x^0}_{t_0} d z^0 \int d^3 z 
&\, \bigl[\Sigma_\rho(x,z)D_F(z,x)-\Sigma_F(x,z)D_\rho(z,x)\nonumber\\
&+D_F(x,z)\Sigma_\rho(z,x)-D_\rho(x,z)\Sigma_F(z,x)\bigr]\,.
\end{align} 
For a spatially homogeneous system $\partial^\mu J_\mu=\partial^0 J_0=\dot{q}(t)$, where $q$ 
is the charge density. Using furthermore \eqref{DFDrhoProp} and   similar relations 
for the self-energies we can simplify \eqref{MasterEquation0} to
\begin{align}
\label{MasterEquation1}
\dot{q}(t)&\equiv S(x)-W(x)  \nonumber\\
&=2\int\limits^{t}_{t_0} d z^0 \int d^3 z\, 
\Im\bigl[\Sigma_\rho(x,z)D_F(z,x)-\Sigma_F(x,z)D_\rho(z,x)\bigr]\,.
\end{align}
This expression gives an exact result for the time derivative of the asymmetry 
(assuming Gaussian initial conditions) and provides the basis for various approximation
schemes, e.g.~the  Boltzmann approximation.

The source and washout terms in \eqref{MasterEquation1} are defined by 
\begin{align}
\label{SourceTerm}
S(x)\equiv 2\int\limits^{x^0}_{t_0} d z^0  \int d^3 z 
\bigl[\,\Im\,\Sigma_\rho(x,z)\Re D_F(z,x) 
 -\Im\,\Sigma_F(x,z)\Re D_\rho(z,x)\bigr]\,,
\end{align} 
and by
\begin{align}
\label{WashoutTerm}
W(x)\equiv -2\int\limits^{x^0}_{t_0} d z^0  \int   d^3 z  
\bigl[\,\Re\,\Sigma_\rho(x,z)\Im D_F(z,x) 
 -\Re\,\Sigma_F(x,z)\Im D_\rho(z,x)\bigr]\,,
\end{align} 
respectively. The definition of the washout term, which should describe the 
washout of a present asymmetry, is motivated by the following considerations. 
The operation of charge conjugation replaces the fields in \eqref{WightmannFunctions} 
by their complex conjugates, see section \ref{FundamentalSymmetries} for more details, 
and the density matrix by the charge conjugate one:
\begin{subequations}
   \label{CCgtrless}
   \begin{align}
   \label{CCgtr}
      D_>(x,y) & \rightarrow  C  D_>(x,y)  C^{-1}
      =\Tr[\mathscr{P}^c\,\bar b(x) b(y)]=D^c_<(y,x)=D^{c\,*}_<(x,y)\,,
      \\
   \label{CCless}
      D_<(x,y) & \rightarrow C D_<(x,y) C^{-1}
      =\Tr[\mathscr{P}^c\,b(y)\bar{b}(x)] = D^c_>(y,x)= D^{c\,*}_>(x,y) \,,
   \end{align}
\end{subequations}
where
we have used relations \eqref{gtrless_properties} in the last equalities of 
\eqref{CCgtr} and \eqref{CCless}. Combining \eqref{WightmannStatSpectr} and 
\eqref{CCgtrless} we find that the \C-conjugated  statistical propagator and 
spectral function are given by
\begin{align}
      D_{F(\rho)}(x,y) \rightarrow C D_{F(\rho)}(x,y) C^{-1}&=D^{c\,*}_{F(\rho)}(x,y)\,.
\end{align}
In a \C-symmetric configuration $\mathscr{P}^c=\mathscr{P}$ and therefore 
the statistical propagator as well as  the spectral function are real-valued in this case.
This implies that, in agreement with physical considerations, the washout 
term, which is proportional to the imaginary part of the propagators, vanishes 
in a \C-symmetric configuration. Let us now consider the source term. To 
this end we need to specify the form of the self-energies. At one-loop level 
they read \cite{Garny:2009qn}
\begin{subequations}
   \label{SigmaTwoLoop}
   \begin{align}
      \Sigma_F(x,y)&=-H^*_{ij} \bigl[\,G_F^{ij}(x,y)D_F(y,x) 
      +{\textstyle\frac14}\,G_\rho^{ij}(x,y)D_\rho(y,x)\bigr]\,,\\
      \Sigma_\rho(x,y)&=+H^*_{ij} \bigl[\,G_F^{ij}(x,y)D_\rho(y,x)
      -\,G_\rho^{ij}(x,y)D_F(y,x)\bigr]\,,
   \end{align}
\end{subequations} 
where we have introduced $H_{ij}\equiv h_{i}h^*_{j}$. The statistical and spectral propagators of 
the mixing  fields are defined analogously to \eqref{StatSpectr}:
\begin{align}
\label{GFGrhoij}
G_{F}^{ij}(x,y)    = {\textstyle\frac12}\langle 
                           \{\psi_i(x),\psi_j(y)\} \rangle\,,\,\,
G_{\rho}^{ij}(x,y) = i\langle [\psi_i(x),\psi_j(y)] \rangle\,.
\end{align}
From the definitions \eqref{GFGrhoij} it follows that
\begin{align}
   \label{GFGrhoProp}
   G_{F}^{ij}(x,y)=G_{F}^{ji}(y,x)\,,\quad
   G_{\rho}^{ij}(x,y)=-G_{\rho}^{ji}(y,x)\,.
\end{align}
Furthermore, using the hermiticity of the density matrix $\mathscr{P}$ and 
the cyclic invariance of the trace, one can show that these matrices are 
real-valued. Therefore, in a \C-symmetric configuration:
\begin{align}
\label{SourceTerm1}
S(x) & \equiv -2\,\Im H_{ij}\int\limits^{x^0}_{t_0} d z^0  \int d^3 z 
\, \bigl[G^{ij}_F(x,z) \Pi_\rho(z,x)-G^{ij}_\rho(x,z) \Pi_F(z,x)\bigr]\,,
\end{align} 
where we introduced 
\begin{subequations}
\label{PiFrho}
\begin{align}
\label{PiF}
\Pi_F(z,x)&\equiv  D^s_F(z,x)\,  D^s_F(z,x)
-{\textstyle\frac14} D^s_\rho(z,x)\, D^s_\rho(z,x)\,,\\
\label{Pirho}
\Pi_\rho(z,x)&\equiv 2 \, D^s_F(z,x)\, D^s_\rho(z,x)\,,
\end{align}
\end{subequations}
for notational convenience and the superscript `$s$' refers to a
\C-symmetric configuration.

For spatially homogeneous systems the two-point functions depend only on the difference of 
the spatial coordinates, $\vec{s}\equiv \vec{x}-\vec{y}$,  and it is convenient to 
introduce their partial Wigner-transforms,
\begin{align}
\label{PartWignerTrafo}
D_{F,\,\rho}(x^0,y^0,\vec{p})&\equiv\int d^3s\, e^{-i\vec{p}\vec{s}}\,D_{F,\,\rho}(x^0,y^0,\vec{s})\,.
\end{align}
The definitions for the self-energies are similar. Substituting \eqref{PartWignerTrafo} into 
\eqref{SourceTerm1} we obtain
\begin{align}
\label{SourceTerm2}
S(t)  = -2\,\Im H_{ij}\int\limits^{t}_{t_0}  d t' \int \frac{d^3 q}{(2\pi)^3}
\, \bigl[G^{ij}_F(t,t',\vec{q}) \Pi_\rho(t',t,\vec{q})-
G^{ij}_\rho(t,t',\vec{q}) \Pi_F(t',t,\vec{q})\bigr]\,. 
\end{align} 
Integrating the source term \eqref{SourceTerm2} over $t$ and using the identity
\begin{align}
\int\limits_{t_0}^{t} dt' \int\limits_{t_0}^{t'} dt''\bigl[f(t',t'')+
f(t'',t')\bigr]=\int\limits_{t_0}^{t} dt' \int\limits_{t_0}^{t} dt'' f(t',t'')\,,
\end{align}
we obtain a `symmetrized' expression for the charge density which would be 
generated in the absence of the wash\-out processes:
\begin{align}
\label{SourceTermIntegratedSym}
q_S(t)  =& -\Im H_{ij}\int\limits^{t}_{t_0}  d t' \int\limits^{t}_{t_0}  d t'' 
\int \frac{d^3 q}{(2\pi)^3}  \nonumber\\
&\times \bigl[G^{ij}_F(t',t'',\vec{q}) \Pi_\rho(t'',t',\vec{q})-
G^{ij}_\rho(t',t'',\vec{q}) \Pi_F(t'',t',\vec{q})\bigr]\,.
\end{align} 
Taking furthermore into account that $\Im\,H_{ii}=0$ and using 
$\Pi_{F(\rho)}(t'',t',\vec{q})=\pm\Pi_{F(\rho)}(t',t'',\vec{q})$ as well
as the properties \eqref{GFGrhoProp} we finally arrive at
\begin{align}
\label{SourceTermIntegratedSym1}
q_S(t)  =& -2\,\Im H_{12}\int\limits^{t}_{t_0}  d t' \int\limits^{t}_{t_0}  d t'' 
\int \frac{d^3 q}{(2\pi)^3} \nonumber\\
&\times \bigl[G^{12}_F(t',t'',\vec{q}) \Pi_\rho(t'',t',\vec{q})-
G^{12}_\rho(t',t'',\vec{q}) \Pi_F(t'',t',\vec{q})\bigr]\,. 
\end{align} 
Equation \eqref{SourceTermIntegratedSym1} provides an exact result for the asymmetry 
in the limit in which washout processes can be neglected. Importantly, it does not 
rely on the quasiparticle approximation and can be used to study the off-shell and 
oscillation effects possibly relevant in the resonant regime.

\paragraph{Equilibrium solution.}
To evaluate  \eqref{SourceTermIntegratedSym1} we need explicit 
expressions for the off-diagonal components of the two-point functions of the 
mixing fields. These are solutions of the corresponding Kadanoff-Baym equations. 
For Gaussian initial conditions the latter take the form \cite{Garny:2009qn}
\begin{subequations}
   \label{KBeqs_real}
   \begin{align}
      \hspace{-1mm}
      \label{GFeq}
      [\square_x+M_{ik}^2]G_{F}^{kj}(x,y) & =
      \int\limits^{y^0}_{t_0} d^4z
      \,\Pi_{F}^{ik}(x,z)G_{\rho}^{kj}(z,y)
      - \int\limits^{x^0}_{t_0} d^4z
      \,\Pi_{\rho}^{ik}(x,z)G_{F}^{kj}(z,y)\,,\\
      \hspace{-1mm}
      \label{Grhoeq}
      [\square_x+M_{ik}^2]G_{\rho}^{kj}(x,y) & =
      \int\limits_{x^0}^{y^0} d^4z\,
      \Pi_{\rho}^{ik}(x,z)G_{\rho}^{kj}(z,y)\,,
   \end{align}
\end{subequations}
where $M_{ij}$ are \emph{mass parameters} of the renormalized Lagrangian
and $\Pi^{ij}_{F,\rho}$ are renormalized self-energies. 
At one-loop level the self-energies are given by \cite{Garny:2009rv,Garny:2009qn} 
\begin{subequations}
\label{PiijFrho}
\begin{align}
\label{PiijF}
\Pi^{ij}_F (x,y)=&-\frac12 H_{ij} 
\bigl[D^2_F(x,y)-{\textstyle\frac14} D^2_\rho(x,y)\bigr]
-\frac12 H^*_{ij} 
\bigl[D^2_F(y,x)-{\textstyle\frac14} D^2_\rho(y,x)\bigr]\,,\\
\label{Pijrho}
\Pi^{ij}_\rho (x,y)=&-\frac12 H_{ij} 
\bigl[2 D_F(x,y)D_\rho(x,y)\bigr]
+\frac12 H^*_{ij} 
\bigl[2 D_F(y,x)D_\rho(y,x)\bigr]\,,
\end{align}
\end{subequations}
Comparing \eqref{PiijFrho} to \eqref{PiFrho} we conclude that in a \C-sym\-me\-tric 
configuration
\begin{align}
\label{PiFrhoCsymm}
\Pi^{ij}_{F(\rho)}(x,y)=-\Re H_{ij}\,\Pi_{F(\rho)}(x,y)\,.
\end{align}
In addition to the statistical and spectral propagators it is also convenient to 
introduce the retarded and advanced ones,
\begin{subequations}
\label{GRGAdef}
\begin{align}
   G_{R}^{ij}(x,y) & \equiv \theta(x^0-y^0)G_{\rho}^{ij}(x,y)\,,\\
   G_{A}^{ij}(x,y) & \equiv -\theta(y^0-x^0)G_{\rho}^{ij}(x,y)\,.
\end{align}
\end{subequations}
The Kadanoff-Baym equations for the retarded and advanced propagators can be derived 
from \eqref{Grhoeq}:
\begin{align}
   \label{GRAQK}
   [\square_x+M_{ik}^2] & G_{R(A)}^{kj}(x,y)=\delta(x-y)\delta^{ij}
                      -\int d^4 z\, \Pi_{R(A)}^{ik}(x,z)G_{R(A)}^{kj}(z,y)\,.
\end{align}
Explicit expressions for the retarded and advanced self-energies can be 
obtained from \eqref{Pijrho}:
\begin{align}
\label{PijRA}
\Pi^{ij}_{R(A)} (x,y)=&-\frac12 H_{ij} 
\bigl[2 D_F(x,y)D_{R(A)}(x,y)\bigr]
-\frac12 H^*_{ij} 
\bigl[2 D_F(y,x)D_{A(R)}(y,x)\bigr]\,.
\end{align}
Since  $D^s_F(x,y)=D^s_F(y,x)$ and  $D^s_R(x,y)=D^s_A(y,x)$ in a \C-symmetric configuration
we conclude that, similarly to \eqref{PiFrhoCsymm}:
\begin{align}
\label{PiRACsymm}
\Pi^{ij}_{R(A)}(x,y)=-\Re H_{ij}\,\Pi_{R(A)}(x,y)\,,
\end{align}
where we have introduced
\begin{align}
\label{PiRA}
\Pi_{R(A)}(x,y)&\equiv 2 \, D^s_F(x,y)\, D^s_{R(A)}(x,y)\,.
\end{align}
Using the definitions of the retarded and advanced propagators \eqref{GRGAdef}, 
we can rewrite the Kadanoff-Baym equations \eqref{KBeqs_real} in the form
\begin{align}
\label{GFrhoeq}
 [\square_x+M_{ik}^2]G_{F(\rho)}^{kj}(x,y)  = &  -\int  d^4z\,\theta(z_0-t_0) \nonumber\\
&\times \bigl[\Pi_{F(\rho)}^{ik}(x,z)G_{A}^{kj}(z,y)
       + \Pi_{R}^{ik}(x,z)G_{F(\rho)}^{kj}(z,y)\bigr]\,.
\end{align}
In thermal equilibrium all two-point functions must be translationally invariant. 
Wigner-transforming the left- and right hand side of \eqref{GRAQK} we obtain,
\begin{align}
   \label{GRAQKWigner}
   \Omega^{ik}_{R(A)}(q)\, G^{kj}_{R(A)}(q)=-\delta_{ij}\,,
\end{align}
where we have introduced 
\begin{align}
\label{OmegaRADef}
\Omega^{ik}_{R(A)}(q)\equiv q^2 \delta_{ik}-M^2_{ik} - \Pi^{ik}_{R(A)}(q)\,.
\end{align}
To reach exact thermal equilibrium the system needs an infinite amount of 
time. Therefore, in order to obtain an equilibrium solution for the statistical
propagator and spectral function we should send the initial time $t_0$ in 
\eqref{GFrhoeq} to minus infinity. Wigner-transforming \eqref{GFrhoeq} and 
using the explicit form of the equilibrium solution for the retarded propagator,
\eqn\eqref{GRAQKWigner}, we obtain
\begin{align}
\label{GFrhoEqSolWigner}
G^{ij}_{F(\rho)}(q)& =-G^{ik}_R(q)\Pi^{kl}_{F(\rho)}(q)G^{lj}_A(q)\,.
\end{align}
Using the Kubo-Martin-Schwinger (KMS) relation one can show that in thermal equilibrium 
the statistical propagator is proportional to the spectral function,
\begin{align}
\label{GFGrhoKMS}
G^{ij}_{F}(q)=\bigl[\textstyle{\frac12}+f(qu)\bigr]\,G^{ij}_{\rho}(q)\,,
\end{align}
where $u$ is the four-velocity of the medium and $f$ is the Bose-Einstein distribution function.
Note that the four-vector $q$ in \eqref{GFGrhoKMS} is not constrained to be on-shell. This implies 
that in equilibrium the spectral shape of the statistical propagator is determined by the 
shape of the spectral function. Let us also note that at one-loop level \eqref{GFGrhoKMS}
also follows from \eqref{PiFrhoCsymm} and \eqref{GFrhoEqSolWigner}, which serves as a 
cross-check of the calculation.

\paragraph{A non-equilibrium solution.}

Using \eqref{GRAQK} one can show that
\begin{align}
\label{GFrhoEqSol}
G^{ij}_{F(\rho)}(x,y)& =-\int^{\infty}_{t_0} d^4 u \int^{\infty}_{t_0} d^4 v
\, G^{ik}_R(x,u)\Pi^{kl}_{F(\rho)}(u,v)G^{lj}_A(v,y)\,,
\end{align}
is a solution of the Kadanoff-Baym equations \eqref{KBeqs_real} for any value of $t_0$. For 
$t_0\rightarrow -\infty$ its Wigner-transform reverts to \eqref{GFrhoEqSolWigner}.

The assumption that the complex field forms a thermal bath makes the one-loop 
self-energies $\Pi_{F(\rho)}$ translationally invariant. The translational 
invariance of the self-energies essentially renders the Kadanoff-Baym equations 
\eqref{KBeqs_real} linear. 
Therefore  a sum of two solutions is also a solution. Motivated by the form of 
\eqref{GFrhoEqSol} we consider \cite{Anisimov:2008dz,Anisimov:2010dk,Anisimov:2010aq,Garny:2011hg} 
\begin{subequations}
\label{DevFromEq}
\begin{align}
\Delta & G^{ij}_\rho(x,y)= 0\,,\\
\Delta & G^{ij}_F(x,y)= - \int d^3u \int d^3v \,
G^{ik}_R(x^0,\vec{x}-\vec{u})\Delta_F^{kl}(\vec{u}-\vec{v})G^{lj}_A(-y^0,\vec{v}-\vec{y})\,.
\end{align}
\end{subequations}
Substituting \eqref{DevFromEq} into \eqref{KBeqs_real} and using \eqref{GRAQK} we 
see that it solves the Kadanoff-Baym equations, except for Dirac-deltas 
located on the initial time surface. In other words, \eqref{DevFromEq} is a weak solution of \eqref{KBeqs_real}.
The delta-functions  can be associated with external 
sources that (instantly) bring the system out of equilibrium at $t=0$. As can be inferred from 
the form of \eqref{DevFromEq}, it is not time-translationally invariant, but 
is space-translationally invariant. Therefore, it is convenient to perform 
the partial Wigner transformation,
\begin{subequations}
\label{DevFromEqWigner}
\begin{align}
\Delta  G^{ij}_\rho & (x^0,y^0,\vec{q})= 0\,,\\
\Delta  G^{ij}_F & (x^0,y^0,\vec{q})= - 
G^{ik}_R(x^0,\vec{q})\Delta_F^{kl}(\vec{q})G^{lj}_A(-y^0,\vec{q})\,.
\end{align}
\end{subequations}
Physically, the sum of the solutions \eqref{GFrhoEqSol} and \eqref{DevFromEq}
can be interpreted as follows. The system of mixing real fields coupled to a 
thermal bath of the complex field begins its evolution at $t_0=-\infty$ in 
a thermal state. At $t=0$ an external source \textit{instantly} brings it out of 
equilibrium. After that it slowly thermalises producing some asymmetry. Because 
the thermal bath remains in equilibrium this asymmetry would eventually be 
completely erased by the washout processes. However, since we neglect the
latter here the asymmetry asymptotically reaches a constant value. 

Of course, the washout processes are physically very important and must be 
taken into account in a phenomenological analysis. We would also like to stress 
that \eqref{DevFromEq} is only applicable for the very peculiar instant external 
perturbation of the system. Nevertheless, even considering this particularly 
simple solution and neglecting the washout effects one can study interesting and 
qualitatively important features of the process of asymmetry generation in the 
regimes which are in principle not accessible in the other methods.

\paragraph{Density matrix and Boltzmann approximations.}

To conclude this section we will demonstrate how one can recover the Boltzmann and 
density matrix approximations for the source term from \eqref{SourceTermIntegratedSym1} 
and provide an interpretation for the solution \eqref{DevFromEqWigner} in terms of 
one-particle distribution functions and the density matrix respectively.  

First of all we send the initial time $t_0$ to minus infinity and express the 
statistical and spectral propagators in terms of the Wightmann two-point functions. 
Then \eqref{SourceTermIntegratedSym1} takes the form 
\begin{align}
\label{SourceTermBoltzmann0}
q_S(t) = & -i\,\Im H_{ij}\int\limits^{t}_{-\infty}  d t' \int\limits^{t}_{-\infty}  d t''  
\, \int \frac{d^3 q}{(2\pi)^3} \frac{d^3 p}{(2\pi)^3}\frac{d^3 k}{(2\pi)^3}\,
(2\pi)^{3}\,\delta(\vec{q}-\vec{p}-\vec{k}) \\
&\times \bigl[G^{ij}_<(t',t'',\vec{q}) D^s_>(t'',t',\vec{p}) D^s_>(t'',t',\vec{k}) 
-G^{ij}_>(t',t'',\vec{q}) D^s_<(t'',t',\vec{p}) D^s_<(t'',t',\vec{k})\bigr]\,.\nonumber
\end{align} 
Next we introduce centre and relative time coordinates, $\tau \equiv (t'+t'')/2$ 
and $s\equiv t'-t''$. The Jacobian determinant of this transformation is unity. Expressed in 
terms of the centre and relative coordinates the integral becomes 
\begin{align}
\label{SourceTermBoltzmann1}
q_S(t) = & -i\,\Im H_{ij}\int\limits^{t}_{-\infty}  d \tau \int\limits^{+\infty}_{-\infty}  d s  
\, \int \frac{d^3 q}{(2\pi)^3} \frac{d^3 p}{(2\pi)^3}\frac{d^3 k}{(2\pi)^3}\,
(2\pi)^{3}\,\delta(\vec{q}-\vec{p}-\vec{k}) \\
&\times \bigl[G^{ij}_<(\tau,s,\vec{q}) D^s_>(\tau,-s,\vec{p}) D^s_>(\tau,-s,\vec{k})
-G^{ij}_>(\tau,s,\vec{q}) D^s_<(\tau,-s,\vec{p}) D^s_<(\tau,-s,\vec{k})\bigr]\,.\nonumber
\end{align}
Introducing a Wigner-transformation with respect to the relative time,
\begin{align}
\label{WignerTransformRelTime}
G^{ij}_{\gtrless}(\tau,s,\vec{q})&=\int^\infty_{-\infty}\frac{dq_0}{2\pi}e^{-iq_0 s} G^{ij}_{\gtrless}(\tau,q_0,\vec{q})\,,
\end{align}
(and a similar definition for $D_\gtrless$) we can rewrite \eqref{SourceTermBoltzmann1} as
\begin{align}
\label{SourceTermBoltzmann2}
&q_S(t) = -i\,\Im H_{ij}\int\limits^{t}_{-\infty}  d \tau
\, \int \frac{d^4 q}{(2\pi)^4} \frac{d^4 p}{(2\pi)^4}\frac{d^4 k}{(2\pi)^4}\,
(2\pi)^4\,\delta(q-p-k)\nonumber\\
&\times \bigl[G^{ij}_<(\tau,q_0,\vec{q}) D^s_>(\tau,p_0,\vec{p}) D^s_>(\tau,k_0,\vec{k})
-G^{ij}_>(\tau,q_0,\vec{q}) D^s_<(\tau,p_0,\vec{p}) D^s_<(\tau,k_0,\vec{k})\bigr]\,. 
\end{align} 
From \eqref{GFGrhoProp} and \eqref{WignerTransformRelTime} it follows that for the mixing real fields:
\begin{align}
G^{ij}_\gtrless(\tau,-q_0,-\vec{q})=G^{ji}_\lessgtr(\tau,q_0,\vec{q})\,.
\end{align}
Similarly, for the complex field in a \C-symmetric configuration:
\begin{align}
D^s_\gtrless(\tau,-p_0,-\vec{p})=D^s_\lessgtr(\tau,p_0,\vec{p})\,.
\end{align}
Using these properties, we can reduce the integrations over positive and 
negative frequencies to integrations over the positive frequencies only:
\begin{align}
\label{SourceTermBoltzmann4}
&q_S(t)  = \,\Im H_{ij} \int\limits^{t}_{-\infty}  d \tau
\int \frac{\theta(q_0) d^4 q}{(2\pi)^4} \frac{\theta(p_0) d^4 p}{(2\pi)^4}\frac{\theta(k_0) d^4 k}{(2\pi)^4}\,
(2\pi)^4\,\delta(q-p-k)\\
&\times \bigl[\Im\, G^{ij}_<(\tau,q_0,\vec{q})\, 2 D^s_>(t,p_0,\vec{p}) D^s_>(t,k_0,\vec{k})
-\Im\, G^{ij}_>(\tau,q_0,\vec{q})\, 2 D^s_<(t,p_0,\vec{p}) D^s_<(t,k_0,\vec{k})\bigr]\,.\nonumber 
\end{align} 
The factors of two in the squared brackets correspond to a sum of the decays 
into particles and antiparticles. 

First we consider the Boltzmann approximation. To introduce a quasiparticle approximation 
for the mixing fields we note that for a \emph{hierarchical mass spectrum} the diagonal 
(in the basis where the mass matrix is diagonal) components of the two-point functions
strongly peak on the corresponding mass shells \cite{Garny:2009qn}. The off-diagonal 
components of the two-point functions are induced by the off-diagonals of the self-energy,
peak at both mass shells and are small, of the order of $\Gamma/\Delta M$. Motivated by 
this observation, we introduce diagonal two-point functions, which are solutions of 
\eqref{GRAQKWigner} and \eqref{GFrhoEqSolWigner} with the off-diagonal components of the
self-energy set to zero \cite{Garny:2009qn}:
\begin{subequations}
\label{Gdiag}
\begin{align}
\label{GRAdiag}
\Omega^{ii}_{R(A)}(q)\mathcal{G}^{ii}_{R(A)}(q)&=-1\,,\\
\label{GFrhodiag}
\mathcal{G}^{ii}_{F(\rho)}(q)& = -\mathcal{G}^{ii}_R(q)\Pi^{ii}_{F(\rho)}(q)\mathcal{G}^{ii}_A(q)\,.
\end{align}
\end{subequations}
The diagonal spectral function strongly peaks on the corresponding mass shell and in the 
limit of vanishing decay width it can be approximated by a delta-function: 
\begin{align}
\label{GrhoDiagQP}
\mathcal{G}^{ii}_\rho(q) & = (2\pi)\,\sign(q_0)\,\delta(q^2-M_i^2)\,.
\end{align}
Similarly to \eqref{GFGrhoKMS}, in equilibrium:
\begin{align}
\label{GFGrhoDiagKMS}
\mathcal{G}^{ii}_{F}(q_0,\vec{q})=\bigl[\textstyle{\frac12}+f_i(qu)\bigr]\,
\mathcal{G}^{ii}_{\rho}(q_0,\vec{q})\,,
\end{align}
Motivated by \eqref{GFGrhoDiagKMS} we use the Kadanoff-Baym ansatz for the diagonal 
two-point functions, i.e.~assume that for small deviations from equilibrium:
\begin{align}
\label{KBAnsatzDiag}
\mathcal{G}^{ii}_{F}(t,q_0,\vec{q})=\bigl[\textstyle{\frac12}+f_i(t,qu)\bigr]\,
\mathcal{G}^{ii}_{\rho}(q_0,\vec{q})\,.
\end{align}
Using the definitions \eqref{Gdiag} we can express the full statistical and spectral
propagators in terms of the diagonal ones. The exact expressions can be found in 
\cite{Garny:2009qn}. Here we will need only the leading-order approximation 
\begin{align}
\label{GgtrlessInTermsOfDiag}
G^{ij}_{F(\rho)}(q)& \approx \delta_{ij}\,\mathcal{G}^{ij}_{F(\rho)}(q)
-(1-\delta_{ij})\bigl[\mathcal{G}^{ii}_{R}(q)\Pi^{ij}_{R}(q)\mathcal{G}^{jj}_{F(\rho)}(q)
+\mathcal{G}^{ii}_{F(\rho)}(q) \Pi^{ij}_{A}(q)\mathcal{G}^{jj}_{A}(q)\bigr]\,,
\end{align}
where no summation over the indices is implied.
Substituting \eqref{GgtrlessInTermsOfDiag} into \eqref{SourceTermBoltzmann4},
using for the diagonal propagators the Kadanoff-Baym ansatz \eqref{KBAnsatzDiag} 
together with the quasiparticle approximation \eqref{GrhoDiagQP}, as well 
as similar approximations for the complex field we obtain
\begin{align}
\label{SourceTermBoltzmann}
q_S(t) & = \sum_i \int\limits^{t}_{-\infty} \! d \tau \!
\int d\Pi^3_q d\Pi^3_p d\Pi^3_k\, (2\pi)^4\,\delta(q-p-k)\,H_{ii} \epsilon_i   \nonumber\\
&\times \bigl\{f_i(t,\vec{q})\, 2 \bigl[1+f^s_b(t,\vec{p})\bigr] \bigl[1+f^s_b(t,\vec{k})\bigr]
-\bigl[1+f_i(t,\vec{q})\bigr]\, 2 f^s_b(t,\vec{p}) f^s_b(t,\vec{k})\bigr\}\,, 
\end{align} 
where $d\Pi^3_q=d^3q/[(2\pi)^3  2\omega_q]$ is the Lorentz-invariant phase-space 
integration measure. The \CP-violating parameters read \cite{Garny:2009qn}:
\begin{align}
   \label{epsilonclassic}
   \epsilon_i = 
   {\rm Im}\biggl(\frac{H_{ij}}{H^*_{ij}}\biggr)
   \frac{(M_i^2-M_j^2)(M_j\Gamma_j L_{\rho})}{(M_i^2 -M_j^2)^2+(M_j \Gamma_j L_\rho)^2}
   \approx \epsilon_i^{vac} L_\rho\,,
\end{align}
where $\Gamma_j=H_{jj}/(16\pi M_j)$ is the tree-level decay width and we have neglected
the momentum-de\-pe\-n\-de\-nce of the denominator to obtain the second approximate equality. 
Note that $\epsilon_i$ vanish if either $\Im\, H_{12}=0$, $\Re\,H_{12}=0$ or $M^2_2=M^2_1$. 
This reflects basic \CP-properties of the Lagrangian which we will discuss in more details in section \ref{FundamentalSymmetries}.
The  function $L_\rho$ introduced in \eqref{LrhoDef} takes into account medium corrections  and approaches unity at zero temperature.

Because we assume that the complex field forms a thermal bath with a constant temperature
the one-particle distribution functions $f^s_b$ are time-independent. The one-particle distribution functions of the real fields can be represented  as a sum 
of the equilibrium one and a deviation from equilibrium, $f_i=f_i^{eq}+\Delta f_i$. 
In agreement with the third Sakharov condition the contribution 
of the equilibrium part to the right-hand side of \eqref{SourceTermBoltzmann} is identically zero. The contribution 
induced by the deviation from equilibrium reads 
\begin{align}
\label{SourceTermBoltzmannDevFromEq}
q_S(t) = \sum_i  & \int\limits^{t}_{-\infty} \! d \tau \!
\int  d\Pi^3_q\,2\, \epsilon_i\, H_{ii}\, \Delta f_i(t,\vec{q})\nonumber\\
&\times \int d\Pi^3_p d\Pi^3_k\, (2\pi)^4\,\delta(q-p-k)\, \bigl[1+f^s_b(\vec{p})+f^s_b(\vec{k})\bigr]\,.
\end{align} 
The second line of \eqref{SourceTermBoltzmannDevFromEq} is nothing but a full Wigner 
transform of \eqref{Pirho} in the quasiparticle approximation.

To interpret \eqref{DevFromEqWigner} in terms of one-particle distribution functions 
$\Delta f_i$ we Wigner-transform it with respect to the relative time,
\begin{align}
\label{GFWignerTrafo}
\Delta G^{ij}_F(\tau,q_0,\vec{q})=-2\Delta^{kl}_F(\vec{q})\,\int^\infty_{-\infty} \frac{dp_0}{2\pi} G_R^{ik}(q_0+p_0,\vec{q})G_A^{lj}(q_0-p_0,\vec{q})\,\theta(\tau)e^{-2ip_0\tau}\,.
\end{align}
Because \eqref{DevFromEqWigner} vanishes for $x^0<0$ and $y^0<0$, the Wigner transform 
vanishes for central times $\tau<0$, as is reflected by the $\theta(\tau)$. For $j=i$
in the Boltzmann approximation we can furthermore neglecting the off-diagonal components 
of the propagators as well as off-diagonal components of the matrix $\Delta_F^{kl}$. 
In this case we find 
\begin{subequations}
\label{DeltaFsimplifiedNew}
\begin{align}
\label{OscillatingFunc}
\Delta G^{ii}_F(\tau,q_0,\vec{q})&\approx \frac{\sin[2(q_0-\omega_i)\tau]}{q_0(q_0-\omega_i)}\cdot\Delta f_{i}(\tau,\vec{q})\,,\\
\label{Deltafapprox}
\Delta f_{i}(\tau,\vec{q})&\equiv -\frac{\Delta_F^{ii}(\vec{q})}{2\omega_i}\theta(\tau)e^{-\Gamma_i\tau}\equiv 
\Delta f_{i}(0,\vec{q})\theta(\tau)e^{-\Gamma_i\tau}\,.
\end{align}
\end{subequations}
For $\omega_i\tau\gg 1$ the first factor in \eqref{OscillatingFunc} strongly peaks in 
the vicinity of the mass shell, $q_0\sim \omega_i$, and rapidly oscillates away from 
the mass shell. The integration in the proximity of $q_0=\omega_i$ yields a result 
which oscillates around $\frac12$ with amplitude which decreases for increasing $\tau$.
In other words, effectively,
\begin{align}
\label{DeltaFsimplified}
\Delta G^{ii}_F(\tau,q_0,\vec{q})\approx \Delta f_{i}(\tau,\vec{q})\,G^{ii}_\rho(q_0,\vec{q})\,
\end{align}
for $\omega_i\tau\gg 1$. Comparing \eqref{DeltaFsimplified} to \eqref{KBAnsatzDiag} we 
conclude that $\Delta f_{i}(\tau,\vec{q})$ is a one-particle distribution function. This
implies that in the used approximation $\Delta_F^{ii}(\vec{q})$ parametrizes the initial 
deviation of the one-particle distribution function of $\psi_i$ from the equilibrium 
one (which is determined by the temperature of the thermal bath). Substituting 
\eqref{Deltafapprox} into \eqref{SourceTermBoltzmannDevFromEq}, integrating over time 
and reordering the terms we obtain:
\begin{align}
\label{qsBoltzmann}
q_S(t) & =2\sum_i\left(1-e^{-\Gamma_i t}\right) 
\int \frac{dq^3}{(2\pi)^3}\frac{M_i}{\omega_i}\, \Delta f_i(0,\vec{q})\epsilon_i(\omega_i,\vec{q}) 
L_\rho(\omega_i,\vec{q})\,.
\end{align} 
Note that in the small width limit, even though $\omega_i\tau\gg 1$ is crucial for the applicability of the approximation, we can still assume $\Gamma \tau\ll 1$ at the lower limit of the time integration. 
Expression \eqref{qsBoltzmann} has a simple physical interpretation. In the absence of washout 
processes the final asymmetry is expected to be proportional to a product of the initial 
deviation from equilibrium, $\Delta f_i(0,\vec{q})$, and the in-medium \CP-violating
parameter, $\epsilon_i$. The overall factor of two in \eqref{qsBoltzmann} reflects 
the fact that in the toy model considered here `lepton' number is violated by two 
units in each decay. The factor $L_\rho$ comes from the difference of the gain and loss 
terms.It is sometimes interpreted as a medium correction to the decay width, $\Gamma^{med}_i=
\Gamma_i\,L_\rho$. However, we would like to stress that the so defined effective width 
does not coincide with the effective width inferred from the analysis of the spectral 
function. Finally, the Lorentz-invariant integration measure coincides with the one 
that arises in the calculation of the decay reaction density. For comparison with the 
results of the first-principles approach a well as those obtained in the density matrix approximation discussed below it is also useful to rewrite \eqref{qsBoltzmann} in the form
\begin{align}
\label{BlzmnAsymmAppr}
q_S(t)\approx\int \frac{d^3q}{(2\pi)^3}\Delta_F(\vec{q})
\frac{-J}{\det M}\,\frac{1}{(M^2_1-M^2_2)^2} 
\,\sum_{i=1,2} \frac{\Pi^2_\rho(\omega_i,\vec{q})}{(2 \omega_i)^2} 
\frac{1-e^{-\Gamma_it}}{\Gamma_i}\,,
\end{align} 
where $J$ is the basis-invariant measure of \C-violation, see equation \eqref{JarlskogInv},
and we have assumed that $\Delta_F^{kl}(\vec{q})=\delta^{kl}\Delta_F(\vec{q})$, see 
section \ref{FundamentalSymmetries} for more details.

Let us now consider the density matrix approximation.
The density matrix \cite{Sigl:1992fn} is related to the Wightmann function by 
\cite{Cirigliano:2009yt}
\begin{align}
\label{DensMatrDef}
\rho^{ij}(t,\vec{q})=\int^\infty_0 \frac{dq_0}{2\pi}\,2q_0\,G^{ij}_<(t,q_0,\vec{q})\,.
\end{align}
As can be inferred from \eqref{DensMatrDef} the mass shells of the Wightmann 
function (see section \ref{TwoRegimes} for more details) are `summed over' in 
\eqref{DensMatrDef}, which is well motivated for a \textit{quasidegenerate} mass spectrum. 
Using a normalization condition for the spectral function \cite{Garny:2009qn}
and taking furthermore into account that in the setup considered here the exact 
spectral function is real-valued we find 
\begin{align}
\label{DensMatrFromGgr}
\int^\infty_0 \frac{dq_0}{2\pi}\,2q_0\,G^{ij}_>(t,q_0,\vec{q})=
\rho^{ij}(t,\vec{q})+\delta^{ij}\,.
\end{align}
Using \eqref{DensMatrDef} and \eqref{DensMatrFromGgr} we find, approximately, from \eqref{SourceTermBoltzmann4}
\begin{align}
\label{SourceTermDensMatr}
& q_S(t) = \Im H_{ij} \int\limits^{t}_{-\infty}  d \tau
\int d\Pi^3_qd\Pi^3_pd\Pi^3_k\,(2\pi)^4\,\delta(q-p-k)\nonumber \\
&\times \Im\bigl\{\rho^{ij}(t,\vec{q})\, 2 \bigl[1+f^s_b(t,\vec{p})\bigr] \bigl[1+f^s_b(t,\vec{k})\bigr]
-\bigl[\delta^{ij}+\rho^{ij}(t,\vec{q})\bigr]\, 2 f^s_b(t,\vec{p}) f^s_b(t,\vec{k})\bigr\}\,.
\end{align} 
Similarly to the Boltzman case, because the equilibrium component of $\rho^{ij}$
does not contribute to the asymmetry, we can simplify the above expression to 
\begin{align}
\label{SourceTermDensMatrDevFromEq}
q_S(t) & = \Im H_{ij} \int\limits^{t}_{-\infty} \! d \tau \!
\int  d\Pi^3_q\,2\, \Im\,\Delta \rho^{ij}(t,\vec{q})\nonumber\\
&\times \int d\Pi^3_p d\Pi^3_k\, (2\pi)^4\,\delta(q-p-k)\, \bigl[1+f^s_b(\vec{p})+f^s_b(\vec{k})\bigr]\,.
\end{align} 
In the small width limit the off-diagonal elements of \eqref{GFWignerTrafo} are 
well approximated (for not too small $q_0$) by 
\begin{align}
\label{DeltaGfWignerQD}
\Delta G^{ij}_F&(\tau,q_0,\vec{q}) \approx
i\Delta_F(\vec{q})\frac{1}{M_i^2-M_j^2}\frac{\Pi_\rho^{ij}(\bar\omega,\vec{q})
}{(2\bar\omega)^2}\nonumber\\
& \times \biggl[
\sum_k\frac{\sin[2(q_0-\omega_k)\tau]}{q_0-\omega_k}\,e^{-\Gamma_k\tau}
-2i\,\frac{\sin[2(q_0-\bar\omega)\tau]}{q_0-\bar\omega}\,
e^{-i(\omega_i-\omega_j)\tau}e^{-\frac12(\Gamma_i+\Gamma_j)\tau}
\biggr]\,,
\end{align}
where $\bar \omega=\frac12(\omega_1+\omega_2)$ and we have again assumed 
that $\Delta_F^{kl}(\vec{q})=\delta^{kl}\Delta_F(\vec{q})$. The $q_0$ integration 
of the first term in the square brackets of \eqref{DeltaGfWignerQD} has been 
discussed above.  Similarly, the integration of the last term in the square 
brackets in the vicinity of $q_0\sim \bar\omega$ gives a term that oscillates around 
$e^{-i(\omega_i-\omega_j)\tau}e^{-\frac12(\Gamma_i+\Gamma_j)\tau}$  with decreasing 
amplitude. Taking this into account, we obtain for the off-diagonals of the 
density matrix: 
\begin{align}
\label{DensMatrAppr}
\Delta\rho^{ij}(\tau,\vec{q}) & \approx
i\Delta_F(\vec{q})\frac{\bar \omega}{M_i^2-M_j^2}\frac{\Pi_\rho^{ij}(\bar\omega,\vec{q})
}{(2\bar\omega)^2}\nonumber\\
& \times \left[e^{-\Gamma_i\tau}+e^{-\Gamma_j\tau} 
-2i\,e^{-i(\omega_i-\omega_j)\tau}e^{-\frac12(\Gamma_i+\Gamma_j)\tau}\right]\,.
\end{align}
Substituting \eqref{DensMatrAppr} into \eqref{SourceTermDensMatr}, using the relation \eqref{PiFrhoCsymm} and definition \eqref{JarlskogInv} we obtain 
for the asymmetry 
\begin{align}
\label{SourceTermDensMatr2}
q_S(t)  & \approx   \int\frac{d^3 q}{(2\pi)^3} 
\Delta_F(\vec{q})\\
&\times \,
\frac{-J}{\det M}\,\frac{1}{(M^2_1-M^2_2)^2}
\frac{\Pi^2_\rho(\bar{\omega},\vec{q})}{(2 \bar{\omega})^2} \Biggl[
\sum_{i=1,2}\frac{1-e^{-\Gamma_it}}{\Gamma_i}
-2\Re \frac{1-e^{-i(\omega_1-\omega_2)t}e^{-\frac12(\Gamma_1+\Gamma_2)t}
}{i(\omega_1-\omega_2)+\frac12 (\Gamma_1+\Gamma_2)}\Biggr]\nonumber\,.
\end{align}
Comparing with \eqref{BlzmnAsymmAppr}, we conclude that the first term in the 
square brackets of \eqref{SourceTermDensMatr2} describes \C-violating decays 
of the heavy particles, whereas the second term describes coherent \C-violating 
oscillations between them which are in principle absent in the Boltzmann 
approximation.

\section{\label{FundamentalSymmetries}Fundamental symmetries and dynamics}

If the system is initially in a \C-symmetric state then a non-zero asymmetry can be generated 
only dynamically. In this section we show that whether the dynamics is \C-conserving or 
\C-violating is determined  by symmetries of the Lagrangian under \C-transformation. For the 
system under consideration the measure of dynamical \C-violation can be parametrized by a single 
flavour-basis invariant combination of the couplings and mass parameters \cite{Hohenegger:2013zia}. On the other hand, 
even if the dynamics is \C-conserving and the initial asymmetry is zero, a non-zero asymmetry 
can be generated provided that the initial conditions for the mixing fields are not \C-symmetric.
Below we discuss the conditions which ensure that the Lagrangian and the initial conditions 
are simultaneously invariant under \C-transformation and show that the obtained results consistently predict zero asymmetry in this case.

\paragraph{Charge conjugation properties.}

If both, the dynamics and the initial conditions, are \C-co\-n\-ser\-ving
then the final asymmetry must be zero. Let us check if \eqref{SourceTermIntegratedSym}
is consistent with this requirement. 

The information about the dynamics is encoded in the Lagrangian. Here 
we work in the \MS-scheme and therefore it is sufficient to analyse only its finite 
part, see \cite{Hohenegger:2013zia} for more details. Under \C-transformation  
the fields transform as,
\begin{subequations}
\label{CTrafos}
\begin{align}
C b(x) C^{-1}&=\beta \bar{b}(x)\,,\\
C \bar{b}(x) C^{-1}&=\beta^* b(x)\,,\\
C \psi_{i}(x) C^{-1}&=U_{ij}\psi_{j}(x)\,.
\end{align}
\end{subequations}
where $\beta$ is a phase factor, $|\beta|=1$, and $U_{ij}$ is an orthogonal 
transformation. The latter can be a flavour rotation or reflection:
\begin{align}
\label{CFlavorTrafos}
U=\left(
\begin{tabular}{cc}
$c$ & $-s$\\
$s$ & $c$
\end{tabular} 
\right)\quad {\rm or}\quad 
U=\left(
\begin{tabular}{cc}
$c$ & $s$\\
$s$ & $-c$
\end{tabular} 
\right)\,,
\end{align}
where we have introduced $c\equiv \cos(\alpha)$ and $s\equiv \sin(\alpha)$ to 
shorten the notation. \C-transforming the Lagrangian \eqref{Lagrangian} and using \eqref{CTrafos}, we find that it
 is \C-invariant provided that
\begin{subequations}
\label{CSymmetryComditions}
\begin{align}
\label{CondM}
U^T_{im}M^2_{mn}U_{nj}&=M^2_{ij}\,,\\
\label{Condh}
\beta^2 U^T_{ik}\, h_{k}&=h^*_{i}\,.
\end{align}
\end{subequations}
If for a given set of couplings and mass parameters, we can 
find $\beta$ and $U_{ij}$ such that the conditions \eqref{CSymmetryComditions} 
are fulfilled then the Lagrangian is \C-invariant.
In general, the mass matrix has non-zero off-diagonal elements. To simplify 
the analysis we rotate to the basis where they vanish, i.e.~where
$M^2_{ij} = \delta_{ij} M_i^2$. If $M_1^2 \neq M_2^2$ 
the first condition is fulfilled only for 
$\alpha=0,\pi$ rotations and $\alpha=0,\pi$ reflections. That is, we have to 
consider only four choices of $U_{ij}$. The condition \eqref{Condh}
is equivalent to the requirement that $U^T_{im}H_{mn}U_{nj}=H^*_{ij}$.
For $\alpha=0,\pi$ rotations this implies $H_{12}=H^*_{12}$ which
 holds if $\Im\,H_{12}=0$. For  $\alpha=0,\pi$ reflections 
\eqref{Condh} implies $H_{12}=-H^*_{12}$, which is fulfilled if 
$\Re\,H_{12}=0$. Under a flavour rotation $\Im\,H_{12}$ and 
$\Re\,H_{12}$ transform as \cite{Hohenegger:2013zia},
\begin{subequations}
\begin{align}
\Im\,H_{12} &\rightarrow \Im\,H_{12}\,,\\
\Re\,H_{12} &\rightarrow (c^2-s^2)\Re\,H_{12}
+cs (H_{22}-H_{11})\,.
\end{align}
\end{subequations}
Therefore, in the special case $M_1^2=M_2^2$ we can always rotate to 
a basis where $\Re\,H_{12}=0$. This implies that the Lagrangian \eqref{Lagrangian} is also 
\C-symmetric in this case. Summarizing the above, the dynamics is 
\C-conserving if either $\Im\,H_{12}=0$, $\Re\,H_{12}=0$ in the basis 
where the mass matrix is diagonal, or the mass matrix is proportional 
to unity. A quantity that vanishes if any of these conditions is fulfilled is
\begin{align}
\label{JarlskogInv}
J=2\,\Im H_{12}\Re H_{12} M_1 M_2 (M^2_2-M^2_1)\,.
\end{align} 
As can readily be checked, this is a special case of the form of $J$ in a general flavour basis:
\begin{align}
J=\Im\,\tr(H M^3 H^T M)\,.
\end{align} 
In other words, $J$ is a basis-invariant measure of \C- and \CP-vi\-o\-la\-tion in the 
theory.
As we showed in \cite{Hohenegger:2013zia} this property is preserved under renormalization.

Let us now analyse in which case the two-point functions are also \C-symmetric. 
As follows from \eqref{CTrafos}, they transform under \C as
\begin{align}
\label{GFCinvariance}
G^{ij}_{F(\rho)}(x,y)& \rightarrow C G^{ij}_{F(\rho)}(x,y) C^{-1} 
=U_{ik}G^{c,kl}_{F(\rho)} (x,y)U^T_{lj}\,.
\end{align}
This is trivially equal to $G^{ij}_{F(\rho)}(x,y)$ for $\alpha=0,\pi$ flavour rotations. 
Thus, if $\Im\,H_{12}=0$, then the source term must vanish irrespective of the 
form of the two-point functions. On the other hand, for $\alpha=0,\pi$
reflections \eqref{GFCinvariance} is invariant only if the two-point functions 
are diagonal in the basis in which the mass matrix is diagonal. Similarly, for 
the special case of a mass matrix proportional to unity the two-point functions 
must be diagonal in the basis in which $\Re H_{12}=0$ to ensure that no asymmetry
is generated. This  implies that vanishing of the invariant \eqref{JarlskogInv}
is in general \textit{insufficient} to ensure zero final asymmetry. Even though the dynamics is
\C-conserving in this case, the asymmetry can be `hidden' in the initial conditions
for the two-point functions of the mixing fields. 

Let us examine if, given $J=0$, the equilibrium solution \eqref{GFrhoEqSol} 
is \C-symmetric. If $\Re H_{12}=0$ in the basis where the mass matrix 
is diagonal then the self-energies are diagonal. Therefore the retarded and 
advanced propagators are flavour-diagonal as well, see \eqref{GRAQKWigner}. This 
in turn implies that the equilibrium spectral and statistical propagators are 
also flavour-di\-a\-gonal, see \eqref{GFrhoEqSolWigner}, and are therefore 
\C-symmetric. If the mass matrix is proportional to unity then, for the same 
reasons, in the basis where $\Re H_{12}=0$ the spectral and statistical 
propagators are diagonal and are again \C-symmetric.

On the other hand, the non-equilibrium part of the solution, \eqn\eqref{DevFromEqWigner}, is not 
necessarily \C-symmetric for $J=0$. Even though the retarded and advanced propagators are 
automatically diagonal if $\Re H_{12}=0$ in the basis where the mass matrix is diagonal,
the resulting matrix $\Delta G^{ij}_F$ is diagonal only if  $\Delta_F^{kl}(\vec{q})$ is also diagonal in this basis. 
If this condition is not fulfilled  then the source term differs from zero even if the 
dynamics is \C-conserving. Similarly, if the mass matrix is proportional to unity then
the non-equilibrium part of the solution is \C-symmetric only provided that 
$\Delta_F^{kl}(\vec{q})$ is diagonal in the basis where $\Re H_{12}=0$.

\paragraph{Contribution of the equilibrium solution.}

Since \eqref{GFrhoEqSol} is the equilibrium solution of \eqref{KBeqs_real}, its contribution 
to the source term is expected to vanish even for $J\neq 0$. Let us check that this is indeed 
the case. First of all, sending, as discussed above, the initial time to minus infinity and 
using the definitions \eqref{GRGAdef} to extend the upper integration limit to plus infinity 
we can rewrite  \eqref{SourceTerm2} in the form
\begin{align}
\label{EquilibriumContribution}
S(t)  & = 2\,\Im H_{ij}\int\limits^{\infty}_{-\infty}  \! d t'  \! 
\int \frac{d^3 q}{(2\pi)^3} \nonumber\\
&\times  \bigl[G^{ij}_F(t-t',\vec{q}) \Pi_A(t'-t,\vec{q}) 
+G^{ij}_R(t-t',\vec{q}) \Pi_F(t'-t,\vec{q})\bigr]\,, 
\end{align} 
where we have taken into account the translational invariance of the equilibrium
solution. Performing a Wigner-transformation with respect to the relative time, 
\begin{subequations}
\label{WignerTransform}
\begin{align}
G^{ij}_{F,\,R,\,A}(\tau,s,\vec{q})&=\int^\infty_{-\infty}\frac{dq_0}{2\pi}e^{-iq_0 s} G^{ij} _{F,\,R,\,A}(\tau,q_0,\vec{q})\,,\\
\label{WignerTransformSpectral}
G^{ij}_{\rho}(\tau,s,\vec{q})&=i\int^\infty_{-\infty}\frac{dq_0}{2\pi}e^{-iq_0 s} G^{ij} _{\rho}(\tau,q_0,\vec{q})\,,
\end{align}
\end{subequations}
we can rewrite it in the form
\begin{align}
\label{EquilibriumContributionWigner}
S&(t)  = 2\,\Im H_{ij} \int \frac{d^4 q}{(2\pi)^4} 
\, \bigl[G^{ij}_F(q_0,\vec{q}) \Pi_A(q_0,\vec{q})
+G^{ij}_R(q_0,\vec{q}) \Pi_F(q_0,\vec{q})\bigr]\,.
\end{align} 
Substituting \eqref{GFrhoEqSolWigner} into \eqref{EquilibriumContributionWigner} and using 
\eqref{PiFrhoCsymm} together with \eqref{PiRACsymm} we can write the contribution of the 
equilibrium solution to the source term as
\begin{align}
\label{EquilibriumContributionWigner1}
S(t) & = 2\,\Im H_{ij} \int \frac{d^4 q}{(2\pi)^4}\,\Pi_F(q_0,\vec{q}) G^{ik}_R(q_0,\vec{q})
\,\bigl[-\Pi^{kl}_A(q_0,\vec{q}) G^{lj}_A(q_0,\vec{q})+\delta_{kj}\bigr]\,.
\end{align} 
Using the solution for the advanced propagator \eqref{GRAQKWigner} we obtain:
\begin{align}
\label{EquilibriumContributionWigner2}
S(t) & = -2\,\Im H_{ij} \int \frac{d^4 q}{(2\pi)^4}\,\Pi_F(q_0,\vec{q})\,
G^{ik}_R(q_0,\vec{q}) \bigl(\,q^2\delta_{kl}-M^2_{kl}\bigr)G^{lj}_A(q_0,\vec{q})\,.
\end{align} 
In a \C-symmetric medium $\Pi_F(-q_0,\vec{q})=\Pi_F(q_0,\vec{q})$. Furthermore, 
$G^{ij}_R(-q_0,\vec{q})=G^{ji}_A(q_0,\vec{q})$. Taking this into account  we 
can rewrite \eqref{EquilibriumContributionWigner2} in the form
\begin{align}
\label{EquilibriumContributionWigner3}
S(t) = & -2\,\Im H_{ij} \int \frac{\theta(q_0) d^4 q}{(2\pi)^4}\,\Pi_F(q_0,\vec{q}) \\
&\times \bigl[ G^{ik}_R(q_0,\vec{q}) \bigl(\,q^2\delta_{kl}-M^2_{kl}\bigr)G^{lj}_A(q_0,\vec{q})
+ G^{jk}_R(q_0,\vec{q}) \bigl(\,q^2\delta_{kl}-M^2_{kl}\bigr)G^{li}_A(q_0,\vec{q})\bigr]=0\,.\nonumber
\end{align} 
Because $\Im H_{ij}$ is antisymmetric with respect to $i\leftrightarrow j$, whereas 
the integrand is symmetric, the contribution of the equilibrium solution vanishes, 
as expected, even for $J\neq 0$.

\paragraph{Contribution of the non-equilibrium solution.}

Next we analyse the contribution of the non-equ\-ili\-bri\-um solution \eqref{DevFromEqWigner}. 
Substituting it into \eqref{SourceTermIntegratedSym1} we obtain
\begin{align}
\label{SourceTermIntegratedNonEq}
q_S(t) & = \Im H_{ij}\int\limits^{t}_{-\infty}  d t' \int\limits^{t}_{-\infty}  d t'' 
\int \frac{d^3 q}{(2\pi)^3} \,
G^{ik}_R(t',\vec{q})\Delta_F^{kl}(\vec{q})G^{lj}_A(-t'',\vec{q}) \Pi_\rho(t'',t',\vec{q})\nonumber\\
&\equiv \int \frac{d^3 q}{(2\pi)^3}\tr\,  \Delta_F(\vec{q})  \eta(\vec{q})\,,
\end{align} 
where we have factored out the dependence on the initial conditions, which are 
encoded in $\Delta_F^{kl}(\vec{q})$, by introducing
\begin{align}
\label{TrB}
\eta^{lk}(t,\vec{q}) & \equiv \Im H_{ij}\int\limits^{t}_{0}  
d t' \int\limits^{t}_{0}  d t'' \,
G^{lj}_A(-t'',\vec{q}) \Pi_\rho(t'',t',\vec{q}) G^{ik}_R(t',\vec{q})\,,
\end{align} 
which contains information on the strength of \C-violation and describes how 
efficiently this initial deviation from equilibrium is converted into 
an asymmetry. Note 
that because of the step functions in the definition of the retarded and advanced propagators 
the lower integration limits in \eqref{TrB} reduce to $t'_{min}=t''_{min}=0$.

Let us now verify that \eqref{SourceTermIntegratedNonEq} disappears if the dynamics
is \C-conserving (i.e.~if $J=0$) and \textit{simultaneously} the initial conditions 
are \C-symmetric. As has been discussed above, if $J=0$ because 
$\Im H_{12}=0$, then the source term must identically vanish. A quick inspection of 
\eqref{TrB} shows that this is indeed the case. If $J=0$ because $\Re H_{12}=0$ 
in the basis where the mass matrix is diagonal, then the self-energy is diagonal. 
As follows from \eqref{GRAQKWigner} the retarded and advanced propagators 
are also diagonal in this case. Since $\Im H_{ii}=0$ this in turn implies 
that the diagonal components of $\eta^{lk}(t,\vec{q})$ become zero. If the non-equilibrium 
solution \eqref{DevFromEqWigner} is \C-symmetric, i.e.~if  $\Delta_F^{kl}(t,\vec{q})$ is 
diagonal in the same basis, then \eqref{SourceTermIntegratedNonEq} automatically 
vanishes, as expected. The case of equal masses is treated analogously.

For a particularly simple choice, $\Delta_F^{kl}(\vec{q})=\delta^{kl} \Delta_F(\vec{q})$, the 
solution \eqref{DevFromEqWigner} is automatically \C-symmetric 
if $J=0$. Therefore, the final asymmetry is expected to be proportional to $J$
and to vanish for $J=0$. On the other hand, one should keep in mind that for this 
choice both the dynamics \textit{and} the initial conditions  are 
\C-violating if $J\neq 0$. Substituting the chosen form of $\Delta_F^{kl}(\vec{q})$ into 
\eqref{SourceTermIntegratedNonEq} we obtain:
\begin{align}
\label{SourceTermIntegratedNonEq1}
q_S(t) & = \int \frac{d^3 q}{(2\pi)^3} \Delta_F(\vec{q})\, \tr\,  \eta(\vec{q})\,.
\end{align} 
Exchanging the integration variables, $t'\leftrightarrow t''$, using the 
property $\Pi_\rho(t'',t',\vec{q})=-\Pi_\rho(t',t'',\vec{q})$ and summing over the 
flavour indices we find:
\begin{align}
\label{TrB1}
\tr\,\eta(t,\vec{q}) & = 2\,\Im H_{12}\int\limits^{t}_{0} d t' \int\limits^{t}_{0}  d t''\, 
\Pi_\rho(t'',t',\vec{q}) \, G^{1n}_R(t',\vec{q}) G^{n2}_A(-t'',\vec{q})\,.
\end{align} 
Replacing the retarded and advanced propagators by their full Wigner-transforms and 
using furthermore that  $G^{ij}_{R(A)}(q_0,\vec{q})= G^{ji}_{R(A)}(q_0,\vec{q})$ 
in a \C-symmetric medium we can rewrite \eqref{TrB1} in the form 
\begin{align}
\label{TrB2}
\tr\,\eta(t,\vec{q}) = & 8\,\Im H_{12}\, \int\limits^{t}_{0} d t' 
\int\limits^{t}_{0}  d t''\, \Pi_\rho(t'',t',\vec{q})
\,\int^\infty_0\frac{dp_0}{2\pi}\int^\infty_0\frac{dk_0}{2\pi}\nonumber\\
&\times \Re\bigl(\bigl[G^{11}_R(p_0,\vec{q})-G^{22}_R(p_0,\vec{q})\bigr]e^{-ip_0t'}\bigr)
\,\Re\bigl(G^{12}_A(k_0,\vec{q})e^{ik_0t''}\bigr)\,.
\end{align} 
Being a trace this expression is flavour-basis invariant and we are free to 
rotate to any other basis. 

\paragraph{Proportionality to the basis-invariant measure of CP-violation.}

We will denote the couplings and masses in the 
new basis by ${\cal H}$ and ${\cal M}$ respectively. As has been argued 
in the previous section, $\Im H_{12}$ is invariant under flavour transformations 
and therefore such a rotation affects only the components of the retarded 
and advanced propagators.
To evaluate  
\eqref{TrB2} we rotate from the basis where the mass matrix is 
diagonal to the basis where $\Re\, {\cal H}_{12}=0$. The rotation angle 
$\alpha$ is given by 
\begin{align}
\label{RotationAngle}
2\alpha= \arctan\left(\frac{2\,\Re H_{12}}{H_{11}-H_{22}}\right)\,.
\end{align}
In the new basis the self-energies are diagonal.  Note that a flavor 
transformation does not `exchange' terms between the basic Lagrangian and the 
counterterms and therefore, in this sense, does not alter the renormalisation 
prescription for the self-energies. Their diagonal components
are proportional to ${\cal H}_{ii}$ which are related to the couplings in 
the initial basis by
\begin{align}
\label{NewCouplings}
{\cal H}_{11/22}& =\frac12 \tr H \pm \frac12 \frac{H_{11}-H_{22}}{\cos 2\alpha}\,.
\end{align}
Using  \eqref{RotationAngle} and \eqref{NewCouplings} we can express $\cos 2\alpha$
and $\sin 2\alpha$ in terms of the couplings in the new and original basis:
\begin{subequations}
\begin{align}
\label{cos2alpha}
\cos 2\alpha=\frac{H_{11}-H_{22}}{{\cal H}_{11}-{\cal H}_{22}}\,,\\
\label{sin2alpha}
\sin 2\alpha=\frac{2 \Re\,H_{12}}{{\cal H}_{11}-{\cal H}_{22}}\,.
\end{align}
\end{subequations}
In the new basis the matrix of mass parameters of the Lagrangian
is no longer diagonal. Its components read
\begin{subequations}
\begin{align}
\label{Msq11and22}
{\cal M}^2_{11/22}&=\frac12\tr\, M^2 \mp \frac12 (M_2^2-M_1^2) \cos 2\alpha\,,\\
\label{Msq12}
{\cal M}^2_{12}={\cal M}^2_{21} & = \frac12 (M_2^2-M_1^2) \sin 2\alpha\,.
\end{align}
\end{subequations}
Using \eqref{GRAQKWigner}, we obtain for the components of the retarded (advanced) 
propagator in the new basis:
\begin{align}
\label{GRANewBasis}
G_{R(A)}=\frac{-1}{\det \Omega_{R(A)}}
\left(
\begin{tabular}{cc}
$\Omega^{22}_{R(A)}$ &  ${\cal M}^2_{12}$\\
${\cal M}^2_{21}$ &  $\Omega^{11}_{R(A)}$
\end{tabular} 
\right)\,.
\end{align}
As follows from \eqref{Msq12} and \eqref{sin2alpha}, the product of $\Im {\cal H}_{12}$ 
and ${\cal M}^2_{12}$ is proportional to the basis-invariant measure of \C-violation: 
\begin{align} 
\label{ImH12M12toJ}
2\, \Im {\cal H}_{12} {\cal M}^2_{12}=\frac{J}{({\cal H}_{11}-{\cal H}_{22}) \det{M}}\,.
\end{align}
Using \eqref{ImH12M12toJ} and the explicit form of the diagonal components of the retarded 
propagator, \eqn\eqref{GRANewBasis},  we can rewrite \eqref{TrB2} in the form 
\begin{align}
\label{TrB3a}
\tr\,\eta(t,\vec{q}) & = \frac{4 J}{\det M}\, \int\limits^{t}_{0} d t' 
\int\limits^{t}_{0}  d t''\, \Pi_\rho(t'',t',\vec{q})\,
\int^\infty_0\frac{dp_0}{2\pi}\int^\infty_0\frac{dk_0}{2\pi}\,
\nonumber\\
&\times \,
\Re\biggl(\frac{\Delta_R(p_0,\vec{q})\,e^{-ip_0t'}}{\det\Omega_R(p_0,\vec{q})}\biggr)
\Re\biggl(\frac{e^{ik_0t''}}{\det \Omega_A(k_0,\vec{q})}\biggr)\,,
\end{align} 
where we have introduced 
\begin{align}
\Delta_R(p_0,\vec{q})&\equiv \frac{(M_2^2-M_1^2)(H_{11}-H_{22})}{(H_{11}-H_{22})^2+(2\Re\,H_{12})^2}+\Pi_R(p_0,\vec{q})\,,
\end{align}
and $\Pi_R(p_0,\vec{q})$ is the Wigner-transform of \eqref{PiRA}. We have 
absorbed the difference ${\cal H}_{11}-{\cal H}_{22}$ in the denominator of 
\eqref{ImH12M12toJ} into the definition of $\Delta_R$ and used \eqref{cos2alpha} 
and \eqref{Msq11and22} to express  $({\cal M}_{22}-{\cal M}_{11})/({\cal H}_{11}-{\cal H}_{22})$
in terms of the couplings and mass parameters in the basis where the mass matrix 
is diagonal. The constant part of $\Delta_R$ is real-valued and, as far as its 
contribution is concerned, can be factored out. The integrand is then symmetric 
under a simultaneous transformation $p_0\leftrightarrow k_0$ and $t'\leftrightarrow t''$.
Therefore, after the integration over $p_0$ and $k_0$ the result is symmetric under 
$t'\leftrightarrow t''$. On the other hand, $\Pi_\rho(t'',t',\vec{q})$ is 
antisymmetric under $t'\leftrightarrow t''$. Thus, the contribution of the constant
part of $\Delta_R$ identically  vanishes as an integral over the product of a symmetric and an
antisymmetric function and we are left with 
\begin{align}
\label{TrB3}
\tr\,\eta(t,\vec{q}) & = \frac{4 J}{\det M}\, \int\limits^{t}_{0} d t' 
\int\limits^{t}_{0}  d t''\, \Pi_\rho(t'',t',\vec{q})\,
\int^\infty_0\frac{dp_0}{2\pi}\int^\infty_0\frac{dk_0}{2\pi}\,
\nonumber\\
&\times \,
\Re\biggl(\frac{\Pi_R(p_0,\vec{q})\,e^{-ip_0t'}}{\det\Omega_R(p_0,\vec{q})}\biggr)
\Re\biggl(\frac{e^{ik_0t''}}{\det \Omega_A(k_0,\vec{q})}\biggr)\,,
\end{align} 
As expected, for \C-symmetric initial conditions the source term is 
proportional to $J$ and vanishes in the absence of dynamical \C-violation. This 
is an important cross check of the self-consistency of the used formalism.

\section{\label{TwoRegimes}\textit{Runaway} and \textit{crossing} regimes}

Whereas thermal effects, by definition, do not affect the couplings and mass 
parameters of the Lagrangian they do influence the effective masses and widths 
of the in-medium quasiparticle excitations. Usually, to minimize the difference 
between the mass parameters and the effective masses and to ensure in this way the 
applicability of the quasiparticle approximation one makes use of the renormalization
group equations and adjusts the renormalization scale for each value of the 
temperature. Effectively, this approach has also been used in our previous 
work \cite{Garny:2011hg}. However, this approach has the disadvantage that the 
relation between the in principle measurable zero-temperature masses and widths 
and the generated asymmetry is not transparent. For this reason here we adopt a 
different approach and fix the renormalization scale at zero temperature. Even 
though the resulting effective masses and widths may substantially deviate from 
the vacuum ones, this poses no technical problems because the Kadanoff-Baym equations 
do not rely on the quasiparticle picture. 

Because the analysis for scalars is technically considerably easier than for 
fermions, here we go beyond the non-relativistic
regime analysed in \cite{Garny:2011hg} and consider also temperatures comparable 
to the masses. We find that, 
depending on the values of the couplings and mass parameters of the Lagrangian, 
the effective masses may either run away from each other, or cross at some point 
as the temperature is increased. The identification of these two regimes is one of the novel aspects
of the present work.

We also demonstrate that for very small mass differences the spectral 
function does not generally peak at the positions corresponding to the effective
masses. This means that the approximation schemes relying on the quasiparticle 
picture (e.g. the Boltzmann equation) are inapplicable in this case. 

\paragraph{Effective masses and widths.}

The right-hand side of \eqref{TrB3} depends on the mass parameters and couplings 
of the Lagrangian \eqref{Lagrangian}. We will consider two benchmark points: 
\begin{subequations}
\begin{align} 
{\rm Set~1:}\quad h_1=0.5\,\mu\,,\quad h_2=0.8\,\mu\exp(2i/3)\,,\quad M_1=\mu\,,\\
{\rm Set~2:}\quad h_1=0.8\,\mu\,,\quad h_2=0.5\,\mu\exp(2i/3)\,,\quad M_1=\mu\,,
\end{align}
\end{subequations}
where $\mu$ is the \MS renormalization scale. The second mass parameter, $M_2$, 
can be expressed in terms of the degeneracy parameter:
\begin{align}
\label{Rdef}
R\equiv \frac{M_2^2-M_1^2}{M_1\Gamma_1+M_2\Gamma_2}\,,
\end{align}
where $M_i\Gamma_i=H_{ii}/(16\pi)$ \cite{Garny:2009qn}.
This definition is motivated by the observation made in \cite{Garny:2011hg} that the 
maximal enhancement of the asymmetry occurs for $M_2^2-M_1^2=M_1\Gamma_1+M_2\Gamma_2$, 
i.e.~for $R\sim 1$. Large $R$ correspond to a hierarchical mass spectrum  and small 
$R$ to a quasi-degenerate mass spectrum. 

Typically, effective masses and widths are defined as real and imaginary parts of the 
zeros of $\det \Omega_R$ (or $\det \Omega_A$, they differ by complex conjugation)
in the complex plane,
\begin{align}
q_{0,I}=\pm\, \omega_{I}-\frac{i}{2} \Geff_{I}\,, \quad
\omega_{I}=(\vec{q}^2+\Meff^2_{I})^\frac12\,.
\end{align}
Note that the effective masses and widths, $\Meff_I$ and $\Geff_I$, are not only 
temperature  but also momentum-dependent because the thermal medium explicitly breaks 
Lorentz-invariance. 

The definition of the effective masses and widths as zeros of $\det \Omega_R$ 
is not completely self-consistent because the retarded self-energy in 
\begin{align}
\label{PoleEquation}
\det\, &\Omega_R=\det\bigl[q^2- M^2-\Pi_R \bigr]
=q^4-q^2\,\tr (M^2+\Pi_R)+\det  (M^2+\Pi_R)\,,
\end{align}
either has to be evaluated for a complex $q_{0,I}$, or the imaginary part of $q_{0,I}$
has to be neglected when evaluating $\Pi_R$. In one-loop approximation the explicit expression 
for the latter reads
\begin{align}
\label{PiRArho}
\Pi_{R,A,\,\rho}(q)=\int \frac{d^4k}{(2\pi)^4} \frac{d^4p}{(2\pi)^4} & (2\pi)^4 \delta(q-p-k)
\,2 D_F^s(p) D^s_{R,A,\,\rho}(k)\,.
\end{align}
It can be represented \cite{Garny:2009rv,Garny:2009qn} as $\Pi_{R,A}=\Pi_{h}\pm \frac{i}2\Pi_\rho$.
In the quasiparticle approximation the Wigner-transforms of the statistical and spectral
propagators of the complex field are given by \cite{Garny:2009rv,Garny:2009qn}
\begin{subequations}
\label{DFrhoQP}
\begin{align}
D^s_\rho(p) & = (2\pi)\,\sign(p_0)\delta(p^2-m^2) \,,\\
D^s_F(p) & = [1+f(pu)]D^s_\rho(p)\,.
\end{align}
\end{subequations}
Using \eqref{DFrhoQP} and assuming $m=0$ we obtain from \eqref{PiRArho}:
\begin{subequations}
\label{PirhoQP}
\begin{align}
\Pi_\rho(q_0,\vec{q})&=\frac{1}{8\pi}L_\rho\left(\frac{q_0}{T},\frac{|\vec{q}|}{T}\right)\,,\\
\label{LrhoDef}
L_\rho(y_0,y)&=1+\frac{2}{y}\ln\left(\frac{1-e^{-(y_0+y)/2}}{1-e^{-(y_0-y)/2}}\right)\,.
\end{align}
\end{subequations}
The retarded and advanced propagators can also be represented in the form $D_{R(A)}=D_h\pm 
\frac{i}{2}D_\rho$\,. In the quasiparticle approximation:
\begin{align}
\label{DhQP}
D^s_h(p) & = -\mathcal{P}\frac1{p^2}\,,
\end{align}
where $\mathcal{P}$ denotes principal value. The resulting dispersive self-energy $\Pi_h$ is divergent 
and must be renormalized.  Using \eqref{DhQP} we find in a \C-symmetric 
configuration:
\begin{subequations}
\begin{align}
\Pi_h(q_0,\vec{q})&=\frac{1}{8\pi}L_h\left(\frac{q_0}{T},\frac{|\vec{q}|}{T}\right)
-\frac{1}{16\pi^2}\ln\frac{|q^2|}{\mu^2}\,,\\
\label{LhDef}
L_h(y_0,y)&=\frac{1}{\pi y}\int\limits_0^\infty dz\, f_{BE}(z)\,
\ln\left|\frac{(2z+y)^2-y_0^2}{(2z-y)^2-y_0^2}\right|,
\end{align}
\end{subequations}
where  $f_{BE}$ denotes the Bose-Einstein distribution
with zero chemical potential. The quasiparticle approximations \eqref{DFrhoQP} and \eqref{DhQP}
are valid only for the real values of their arguments and therefore \eqref{PiRArho} is not well 
defined on the complex plain.
 
To avoid this ambiguity here we use an alternative \emph{self-consistent} definition relying on the 
fact that to an excellent approximation 
\begin{align}
\label{detOmegaRapprox}
\det \Omega_R^{-1}(q_0,\vec{q})\approx\frac{Z}{(q_0^2-q^2_{0,1})(q_0^2-q^2_{0,2})}\,.
\end{align}
Instead of searching for zeros of $\det \Omega_R$ in the complex plane we perform a three-point 
fit which uniquely determines $q_{0,1}$ and $q_{0,2}$ as well as  $Z$. Therefore, even though the 
three fit parameters are complex numbers we do not need to evaluate $\Pi_R$ in the complex plane.
Numerically, the conventional and the alternative definition proposed here give very similar 
results. It is important to note that, because the determinant is invariant under flavour transformations, 
the fit parameters as well as the resulting effective masses and widths do not depend on the choice of 
the flavour basis.

\begin{figure}[t]
      \includegraphics[width=0.49\textwidth]{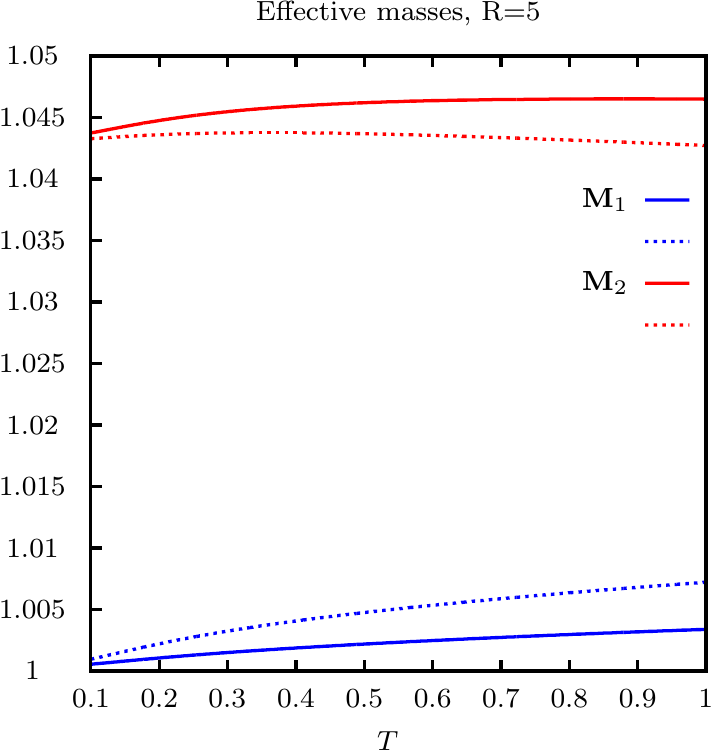} 
      \,\,
      \includegraphics[width=0.5\textwidth]{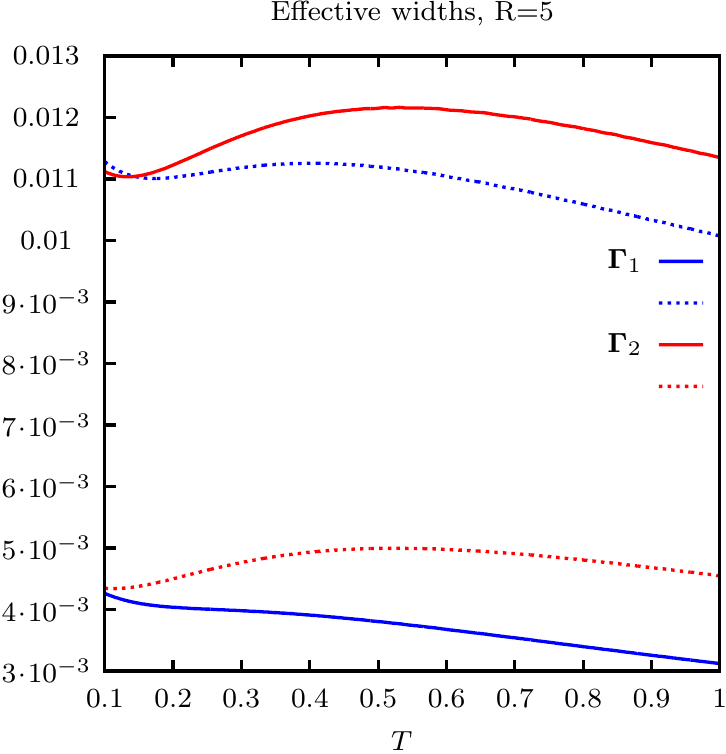}
      \vskip 3mm 
      \includegraphics[width=0.49\textwidth]{figures/EffMassesR1} 
      \,\,
      \includegraphics[width=0.5\textwidth]{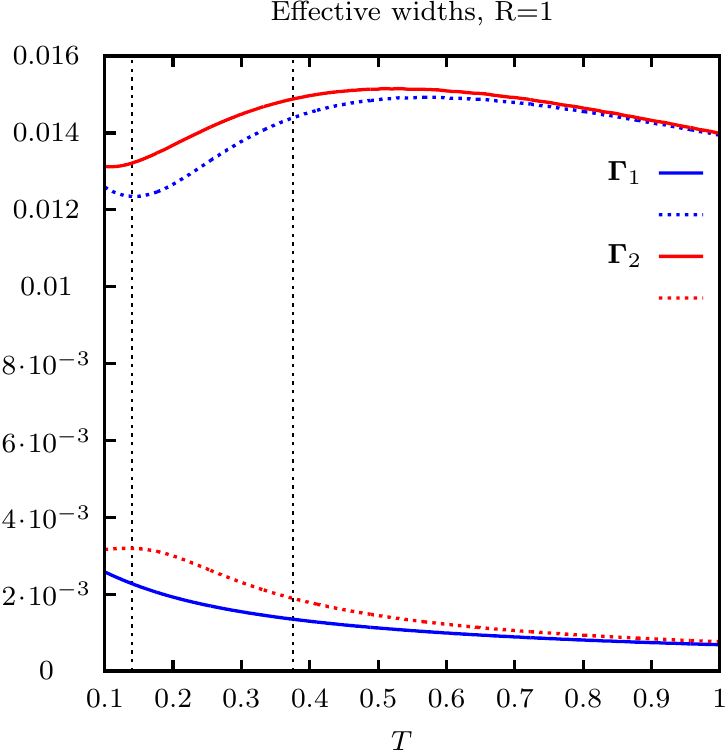}
\caption{\label{EffMassesWidths}Temperature dependence of the effective masses and widths for 
$R=5$ and $R=1$. The solid lines correspond to the first parameter set, and the dotted lines to 
the second parameter set. For $\vec{q}$ we use the average thermal momentum of a boson with 
mass $\mu$ at each $T$. Effective masses, widths and temperature are given in units of $\mu$. The first vertical dotted line indicates the temperature for which the difference between the effective widths takes its minimal value, while the second one indicates the crossing point for which the effective masses are equal.}
\end{figure}

The temperature dependence of the effective masses and widths for the two sets of parameters is 
presented in figure \ref{EffMassesWidths}.
For the first parameter set the difference of the effective masses grows monotonously with the 
temperature. We will refer to this case as to \emph{runaway} regime. It is important to keep in 
mind, that $q_{0,1}$ and $q_{0,2}$ enter \eqref{detOmegaRapprox} symmetrically. Therefore, a transformation
$q_{0,1} \leftrightarrow q_{0,2}$ would `swap' the effective masses, $\Meff_1 \leftrightarrow 
\Meff_2$\,, (as well as the widths) but leave \eqref{detOmegaRapprox} invariant. In figure \ref{EffMassesWidths} 
we have chosen the naming convention for  $\Meff_1$ and $\Meff_2$ such, that in the 
limit $T\rightarrow 0$ these basis-invariant quantities approach the eigenvalues of 
the mass matrix, $M_1$ and $M_2$. This choice is intuitive and convenient, but 
is not forced by any physical principle. For the second parameter set we have also ordered the 
effective masses such that in the limit  $T\rightarrow 0$ they approach $M_1$ and $M_2$ 
respectively. As the temperature grows, the difference of the effective masses decreases and 
at some temperature they become equal. At even higher temperatures the difference of the masses
starts growing again. In principle, we are free to choose the naming convention for the masses 
at any temperature. In other words, we can either assume that one of the effective masses 
continuously grows whereas the other one   continuously decreases (level \emph{crossing}), 
or assume that the effective mass that grew below the crossing temperature, begins to 
decrease, whereas the mass that was increasing before the crossing temperature begins to decrease
(usually referred to as \emph{avoided level crossing}) as the temperature grows further. 
As is evident from figure \ref{EffMassesWidths}  we have chosen the former possibility which 
leads to exactly the same results as the other choice. Note, that due to the momentum dependence 
of the masses the point in which $\Meff_1 = \Meff_2$ is momentum dependent as well. Therefore there 
is no crossing `point' in the strict sense but an interval in which the 
$\Meff_1(\vec{q}) = \Meff_2(\vec{q})$ for typical momenta $\vec{q}$.

At this point we would like to stress once again that \eqref{detOmegaRapprox} is very 
accurate even at the point of (avoided) crossing. In other words the effective masses 
are well defined for any temperature and value of the degeneracy parameter.

We would  also like to emphasize that the definition of 
effective masses and widths used here is not unique and that other definitions are possible. 
To give an example, one could e.g. use $\Pi_h$ in \eqref{PoleEquation} to define the 
effective masses. This definition would lead to a picture of \emph{avoided 
crossing} i.e.~the mass eigenvalues (for a given momentum) would never meet in a point but 
keep a finite minimum distance which would be reminiscent of an avoided crossing with level 
repulsion. The choice of the definition used in this work is motivated mainly by the fact 
that it leads to particularly simple and intuitive expressions for the asymmetry and 
other quantities.

\paragraph{Equilibrium spectral function.}

In the setup considered here the exact spectral function coincides with the equilibrium one. The 
assumption that the spectral function is either diagonal in the basis where the mass matrix 
is diagonal or that its off-diagonal components peak at the same positions as the diagonal 
ones in this basis provides a starting point for various approximation schemes including 
the Boltzmann approximation.
\begin{figure}[t]
	\animategraphics[width=0.5\textwidth,poster=last]{4}{figures/SrhoR5RA/SrhoR5RA}{1}{100}
	\animategraphics[width=0.5\textwidth,poster=last]{4}{figures/SrhoR5CR/SrhoR5CR}{1}{100}
	\vskip 3mm  
	\animategraphics[width=0.5\textwidth,poster=last]{4}{figures/SrhoR1RA/SrhoR1RA}{1}{100} 
	\animategraphics[width=0.5\textwidth,poster=last]{4}{figures/SrhoR1CR/SrhoR1CR}{1}{100}
\caption{\label{SpectralFunction}Components of the equilibrium spectral function for two 
values of the degeneracy parameter and various values of the temperature in the runaway and  
crossing regimes. For $\vec{q}$ we use average thermal momentum of a boson 
with mass $\mu$. To make the comparison with figure \ref{EffMassesWidths} easier 
we have replaced the dependence of the spectral function on $q_0$ by a dependence 
on a mass parameter $M$ using  $q_0\equiv (\vec{q}^2+M^2)^\frac12$. The solid 
vertical lines mark positions the effective masses $\Meff_1$ and $\Meff_2$.
Effective masses and temperature are given in the units of $\mu$. In the PDF-version click on the 
figure to play the animation (supported by Acrobat Reader X).}
\end{figure} 
Components of the spectral function for two values of the degeneracy parameter 
and various values of temperature in the runaway and   crossing regimes 
are presented in figure \ref{SpectralFunction}. To make the comparison with 
figure \ref{EffMassesWidths} easier we have replaced the dependence of the 
spectral function on $q_0$ by a dependence on a mass parameter $M$ defined by  
$q_0\equiv (\vec{q}^2+M^2)^\frac12$.

Given the introduced rescaling one would expect that the diagonal components of 
the spectral function peak at the values of $M$ equal to the values of the two 
effective masses, $\Meff_1$ and $\Meff_2$. Whereas this is the case for 
a hierarchical ($R\gtrsim 5$) and mildly quasi-degenerate ($1\lesssim R\lesssim 5$) mass spectrum, we 
observe a rather different behaviour for a quasi-degenerate ($R\lesssim 1$) mass spectrum,
especially at high temperatures. In particular, in the runway regime at $T\sim \mu$
both diagonal components of the spectral function peak in the vicinity of 
$\Meff_1$ and do not display any non-trivial  features in the vicinity of 
$\Meff_2$, see figure \ref{SpectralFunction}. In the   crossing regime 
we observe a similar behaviour. 

Furthermore, the off-diagonal components differ from zero at any temperature and 
value of the degeneracy parameter. Note that even for a  mildly hierarchical mass 
spectrum the peaks of the off-diagonals are not very pronounced and their positions 
do not exactly coincide with those of the diagonal components. For a quasi-degenerate 
mass spectrum the clear peak structure typical for a hierarchical mass spectrum 
disappears completely. In this case one of the diagonals does not have a pronounced 
peak and the off-diagonal components peak at only one of the mass shells. This makes 
the quasiparticle approximation and the use of the Boltzmann equation in this regime 
rather questionable.

\section{\label{BreitWigner}Analytical treatment of leading effects}

The dominant contribution to the asymmetry is generated by the `difference of frequencies' 
terms close to the mass shell. In this section we evaluate this term analytically in 
the Breit-Wigner approximation. For a hierarchical mass spectrum our first-principle 
approach reproduces the Boltzmann approximation. We would like to stress that, in contrast
to what has been contemplated in other works, in order to obtain the Boltzmann result
there is no need to introduce non-zero widths of the particles forming the thermal bath. 
For a quasi-degenerate mass spectrum the contributions describing destructive interference 
between the two mass shells become important. This renders the Boltzmann approximation 
invalid. In particular, in the crossing regime the asymmetry computed neglecting these 
contributions develops a spurious  peak at temperatures close to the crossing point. The 
negative interference terms `remove' this peak and smoothen the dependence 
of the asymmetry on the temperature.

\paragraph{Double-time integration.}

To evaluate \eqref{TrB3} we replace the spectral `self-energy' $\Pi_\rho$
by its Wigner-transform using a relation similar to \eqref{WignerTransformSpectral}.
The double-time integration can then be performed analytically, e.g.: 
\begin{align}
\label{Fdef}
F(q_0,p_0,k_0,t)&\equiv \int\limits^{t}_{0} d t' \int\limits^{t}_{0}  d t''\, e^{-i q_0(t''-t')}
e^{-ip_0t'}e^{ik_0t''}
=\frac{1-e^{i(q_0-p_0)t}}{q_0-p_0} \frac{1-e^{-i(q_0-k_0)t}}{q_0-k_0}\,.
\end{align}
As can readily be checked, even though the denominator of \eqref{Fdef} vanishes for 
$q_0= p_0$ and $q_0= k_0$,  the numerator simultaneously vanishes as well and the
ratios are finite. Rearranging the terms, we can then rewrite \eqref{TrB3} in the form
\begin{align}
\label{TrB4}
\tr\,\eta(t,\vec{q})= & -\frac{2 J}{\det M}\,
\int^\infty_0\frac{dq_0}{2\pi} 
\int^\infty_0\frac{dp_0}{2\pi}\int^\infty_0\frac{dk_0}{2\pi}\,\,\Pi_\rho(q_0,\vec{q})
\nonumber\\
&\times \,
\Im\biggl(\frac{\Pi_R(p_0,\vec{q})F(q_0,p_0,k_0,t)}{\det\Omega_R(p_0,\vec{q})\det \Omega_A(k_0,\vec{q})}
-\frac{\Pi_R(p_0,\vec{q})F(-q_0,p_0,k_0,t)}{\det\Omega_R(p_0,\vec{q})\det \Omega_A(k_0,\vec{q})}\nonumber\\
&+\frac{\Pi_R(p_0,\vec{q})F(q_0,p_0,-k_0,t)}{\det\Omega_R(p_0,\vec{q})\det \Omega_R(k_0,\vec{q})}
-\frac{\Pi_R(p_0,\vec{q})F(-q_0,p_0,-k_0,t)}{\det\Omega_R(p_0,\vec{q})\det \Omega_R(k_0,\vec{q})}\biggr)\,.
\end{align} 
Because the integration in \eqref{TrB4} is over positive values of $q_0$, $p_0$ and 
$k_0$, the dominant contribution is due to the first term where $q_0-p_0$ and 
$q_0-k_0$ in the denominator can vanish simultaneously. In this section we will 
only consider this `difference of frequencies' contribution and neglect the other 
three. They are studied numerically in section \ref{Numerics}.

To evaluate the remaining momentum integrals we use the approximation \eqref{detOmegaRapprox}. 
Introducing  $x_I\equiv q^2_{0,I}-\vec{q}^2$ we can rewrite its right-hand side in the form, 
\begin{align}
\label{detOmegaRdec}
\frac1{\det\, \Omega_R(q_0,\vec{q})}&\approx 
-\frac{Z}{x_1-x_2}\frac{1}{(q_0^2-q_{0,1}^2)} +\frac{Z}{x_1-x_2}\frac{1}{(q_0^2-q_{0,2}^2)}\,.
\end{align}
Similarly,
\begin{align}
\label{detOmegaAdec}
\frac1{\det\, \Omega_A(q_0,\vec{q})}&\approx 
-\frac{Z^*}{x^*_1-x^*_2}\frac{1}{(q_0^2-q_{0,1}^2)^*} -\frac{Z^*}{x^*_1-x^*_2}\frac{1}{(q_0^2-q_{0,2}^2)^*}\,.
\end{align}
Substituting these expressions, we find for the first term of \eqref{TrB4}:
\begin{align}
\label{TrBEffMGammaAppr}
\tr\,\eta(t,&\,\vec{q})=  -\frac{2 J}{\det M}\,\frac{|Z|^2}{|x_1-x_2|^2}
\int^\infty_0\frac{dq_0}{2\pi}\, \Pi_\rho(q_0,\vec{q})
\\
&\times\Im\biggl[\,
\sum_{I=1,2} \int^\infty_0\frac{dp_0}{2\pi}\frac{\Pi_R(p_0,\vec{q})}{p_0^2-q^2_{0,I}} \frac{1-e^{i(q_0-p_0)t}}{q_0-p_0}
\int^\infty_0\frac{dk_0}{2\pi}\frac{1}{(k_0^2-q^2_{0,I})^*}\frac{1-e^{-i(q_0-k_0)t}}{q_0-k_0}\nonumber\\
&-\sum_{I=1,2}^{I\neq J}\int^\infty_0\frac{dp_0}{2\pi}\frac{\Pi_R(p_0,\vec{q})}{p_0^2-q^2_{0,I}} \frac{1-e^{i(q_0-p_0)t}}{q_0-p_0}
\int^\infty_0\frac{dk_0}{2\pi}\frac{1}{(k_0^2-q^2_{0,J})^*}\frac{1-e^{-i(q_0-k_0)t}}{q_0-k_0}\biggr]\,.\nonumber
\end{align}
Note that the two terms in square brackets have opposite sign, i.e.~there is a destructive 
interference. This effect can be traced back to equations \eqref{detOmegaRdec} and \eqref{detOmegaAdec}.
In the limit $q_{0,2}\rightarrow q_{0,1}$ the difference $x_1-x_2$ in the denominator of \eqref{TrBEffMGammaAppr}
vanishes. However, because of the destructive interference, the numerator simultaneously vanishes as 
well and the ratio remains finite. This conclusion is consistent with the results of \cite{Garny:2011hg}.

\paragraph{Hierarchical mass spectrum.}

For a hierarchical mass spectrum $q_{0,1}$ and $q_{0,2}$ are well separated and the 
contribution of the   interference term is negligible. Therefore, it is sufficient to 
consider only the first term of \eqref{TrBEffMGammaAppr}. Because the integrands are strongly
peaked in the vicinity of $q_{0,I}$, we can approximate $\Pi_R(p_0,\vec{q})$ by 
$\Pi_R(\omega_I,\vec{q})$. The $p_0$ and $k_0$ integrals are then complex conjugates 
of each other and their product is real valued. Since $\Im\, \Pi_R(\omega_I,\vec{q})
=\frac12 \Pi_\rho(\omega_I,\vec{q})$ this results in 
\begin{align}
\label{TrBHierarchical}
\tr\,\eta(t,\vec{q})\approx & -\frac{J}{\det M}\,\frac{|Z|^2}{|x_1-x_2|^2}
\sum_{I=1,2} \Pi_\rho(\omega_I,\vec{q})
\int^\infty_0\frac{dq_0}{2\pi}\, \Pi_\rho(q_0,\vec{q})
\nonumber\\
&\times\,
\int^\infty_0\frac{dp_0}{2\pi}\frac{1}{p_0^2-q^2_{0,I}} \frac{1-e^{i(q_0-p_0)t}}{q_0-p_0}
\int^\infty_0\frac{dk_0}{2\pi}\frac{1}{(k_0^2-q^2_{0,I})^*}\frac{1-e^{-i(q_0-k_0)t}}{q_0-k_0}\,.
\end{align}
The integrations over $p_0$ and $k_0$ give approximately:
\begin{subequations}
\label{Approxp0k0Int}
\begin{align}
\label{Approxp0Int}
\int^\infty_0  \frac{dp_0}{2\pi} \frac1{(p^2_0-q^2_{0,I})} 
\frac{1-e^{i(q_0-p_0)t}}{q_0-p_0}&\approx -\frac{i}{2 q_{0,I}}\frac{1-e^{i(q_0-q_{0,I})t}}{q_0-q_{0,I}}\,,\\
\label{Approxk0Int}
\int^\infty_0  \frac{dk_0}{2\pi} \frac1{(k^2_0-q^2_{0,I})^*} 
\frac{1-e^{-i(q_0-p_0)t}}{q_0-k_0}&\approx \frac{i}{2 q^*_{I,0}}\
\frac{1-e^{-i(q_0-q^*_{I,0})t}}{q_0-q^*_{I,0}}\,.
\end{align}
\end{subequations}
These expressions are valid in the vicinity of $q_0\approx \omega_I$ and result in 
\begin{align}
\label{TrBHierarchical1}
\tr\,\eta(t,\vec{q})\approx & -\frac{J}{\det M}\,\frac{|Z|^2}{|x_1-x_2|^2}
\sum_{I=1,2} \Pi_\rho(\omega_I,\vec{q})
\int^\infty_0\frac{dq_0}{2\pi}\, \Pi_\rho(q_0,\vec{q})
\nonumber\\
&\times \,\frac{1}{(2 \omega_I)^2} 
\frac{|1-e^{i(q_0-\omega_{I})t}e^{-\frac12\Geff_I t}|^2}{
(q_0-\omega_I)^2+(\frac12 \Geff_I)^2}\,.
\end{align} 
Using
\begin{align}
\lim_{\epsilon\rightarrow 0} \frac{2\epsilon}{\omega^2+\epsilon^2}=2\pi\delta(\omega)\,,
\end{align}
we can perform the integration over $q_0$ and obtain 
\begin{align}
\label{TrBhierarchicalappr}
\tr\,\eta(t,\vec{q}) &  \approx -\frac{J}{\det M}\,\frac{|Z|^2}{|x_1-x_2|^2} 
\,\sum_{I=1,2} \frac{\Pi^2_\rho(\omega_I,\vec{q})}{(2 \omega_I)^2} 
\frac{\bigl(1-e^{-\frac12\Geff_It}\bigr)^2}{\Geff_I}\,.
\end{align} 
The obtained time dependence, $(1-e^{-\frac12\Geff_It})^2$, is consistent 
with the result of \cite{Anisimov:2010aq}. However, numerical analysis shows that 
this approximation is rather crude. The oscillating exponent $e^{i(q_0-\omega_I)t}$
substantially changes the shape of the peak in the vicinity of $q_0\approx \omega_I$
and renders this approximation inaccurate. On the other hand, a very accurate 
approximation is provided by 
\begin{align}
\label{TrBhierarchicalacc}
\tr\,\eta(t,\vec{q}) &  \approx -\frac{J}{\det M}\,\frac{|Z|^2}{|x_1-x_2|^2} 
\,\sum_{I=1,2} \frac{\Pi^2_\rho(\omega_I,\vec{q})}{(2 \omega_I)^2} 
\frac{1-e^{-\Geff_It}}{\Geff_I}\,,
\end{align} 
compare with \eqref{BlzmnAsymmAppr}. 
In other words, by taking into account the oscillating exponent we recover the 
time dependence expected in the Boltzmann approximation, see \cite{Anisimov:2010aq,Garny:2011hg},
without introducing an effective width of the particles forming the thermal bath. Note that 
\begin{align}
\label{ResonantEnhancement}
\frac{1}{|x_1-x_2|^2} \approx \frac{1}{(\Meff^2_1-\Meff^2_2)^2+
(\omega_1\Geff_1-\omega_2\Geff_2)^2}\,,
\end{align}
i.e we recover the usual resonant enhancement \cite{Anisimov:2005hr,Garny:2011hg}.
Furthermore, combining \eqref{ResonantEnhancement} with the definition $J$, see 
\eqref{JarlskogInv}, we find that the product of the first two terms in 
\eqref{TrBhierarchicalacc} is equal to,
\begin{align}
\label{EpsilonMathias}
-\frac{J}{\det M}\,\frac{1}{|x_1-x_2|^2} =
\frac{\Im(H^2_{12}) (M_2^2-M_1^2)}{(\Meff^2_1-\Meff^2_2)^2+
(\omega_1\Geff_1-\omega_2\Geff_2)^2}\,,
\end{align}
which strongly resembles equation (137) of \cite{Garny:2011hg} found there empirically using 
numerical analysis. The peculiarity of this expression is that its numerator contains the mass 
parameters of the Lagrangian, whereas the denominator contains effective masses and widths.

Let us finally compare \eqref{TrBhierarchicalacc} with \eqref{qsBoltzmann}. For 
a strongly hierarchical mass spectrum the deviation of the thermal masses and widths from the vacuum 
ones is negligible. Furthermore, the $M_i\Gamma_i$ terms are small compared to $M_i^2-M_j^2$.
Therefore \eqref{EpsilonMathias} is approximately equal to $(16\pi)^2\,\epsilon^{vac}_iM_i\Gamma_i$.
Taking into account that $\Pi_\rho=L_\rho/(8\pi)$ and using \eqref{epsilonclassic} we obtain, 
\begin{align}
\label{tretahhierarchical}
\tr\,\eta(t,\vec{q})&=2 \sum_i (1-e^{-\Gamma_it}) 
\frac1{2\omega_i} \frac{M_i}{\omega_i}\, \epsilon_i(\omega_i,\vec{q}) L_\rho(\omega_i,\vec{q})\,.
\end{align}
In other words, $\eta$ can be viewed as a weighted average of the in-medium \CP-violating parameter
with some kinematical and thermal functions. Substituting \eqref{tretahhierarchical} into 
\eqref{SourceTermIntegratedNonEq1} and recalling the relation between $\Delta_F(\vec{q})$ and 
$\Delta f(0,\vec{q})$ we see that the resulting expression for $q_S(t)$ is identical to  
\eqref{qsBoltzmann}.

\paragraph{Quasidegenerate mass spectrum.}

For a quasi-degenerate mass spectrum $q_{0,1}$ and $q_{0,2}$ are very close and we need to take 
into account also the contributions of the $J\neq I$ term in \eqref{TrBEffMGammaAppr}. 
Since away of $p^2_0=\vec{q}^2$ region  $\Pi_R(p_0,\vec{q})$ is a slowly varying function of $p_0$ we can  approximate it by 
$\Pi_R(\bar{\omega},\vec{q})$, where $\bar{\omega}\equiv \frac12 (\omega_1+\omega_2)$. 
In this approximation the products ($J=I$ contributions) and the sum of the products 
($J\neq I$ contributions) of the momentum integrals in \eqref{TrBEffMGammaAppr} are again 
real valued and we obtain 
\begin{align}
\label{TrBQuasiDegAnalyt1}
\tr\,\eta(t,&\,\vec{q})=  -\frac{J}{\det M}\,\frac{|Z|^2}{|x_1-x_2|^2}\,\Pi_\rho(\bar{\omega},\vec{q})
\int^\infty_0\frac{dq_0}{2\pi}\, \Pi_\rho(q_0,\vec{q}) 
\nonumber\\
&\times\biggl[\,
\sum_{I=1,2} \int^\infty_0\frac{dp_0}{2\pi}\frac{1}{p_0^2-q^2_{0,I}} \frac{1-e^{i(q_0-p_0)t}}{q_0-p_0}
\int^\infty_0\frac{dk_0}{2\pi}\frac{1}{(k_0^2-q^2_{0,I})^*}\frac{1-e^{-i(q_0-k_0)t}}{q_0-k_0}\nonumber\\
&-2\Re \int^\infty_0\frac{dp_0}{2\pi}\frac{1}{p_0^2-q^2_{0,1}} \frac{1-e^{i(q_0-p_0)t}}{q_0-p_0}
\int^\infty_0\frac{dk_0}{2\pi}\frac{1}{(k_0^2-q^2_{0,2})^*}\frac{1-e^{-i(q_0-k_0)t}}{q_0-k_0}\biggr]\,.
\end{align}
Using approximations \eqref{Approxp0k0Int} and integrating over $q_0$ we arrive at 
\begin{align}
\label{TrBQuasiDegAnalyt}
\tr\,\eta(t,\vec{q})   \approx  & -\frac{J}{\det M}\,\frac{|Z|^2}{|x_1-x_2|^2}
\frac{\Pi^2_\rho(\bar{\omega},\vec{q})}{(2 \bar{\omega})^2} 
\nonumber\\
&\times \,\Biggl[
\sum_{I=1,2}\frac{1-e^{-\Geff_It}}{\Geff_I}
-2\Re \frac{1-e^{-i(\omega_1-\omega_2)t}e^{-\frac12(\Geff_1+\Geff_2)t}
}{i(\omega_1-\omega_2)+\frac12 (\Geff_1+\Geff_2)}\Biggr]\,,
\end{align} 
which is analogous to the result of \cite{Garny:2011hg}, compare also with 
\eqref{SourceTermDensMatr2}. As has been discussed above, if  
$\Meff_2\rightarrow \Meff_1$ and $\Geff_2\rightarrow \Geff_1$ simultaneously, then both the 
numerator and the difference $x_1-x_2$ in the denominator of \eqref{TrBQuasiDegAnalyt}
vanish simultaneously, but their ratio remains finite.

\begin{figure}
   \includegraphics[width=0.49\textwidth]{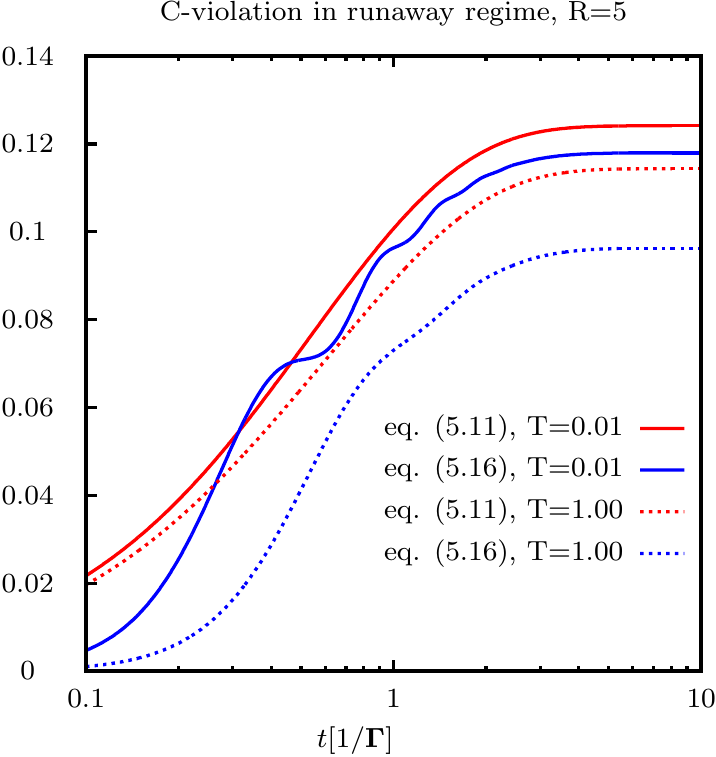}
   \,\,
   \includegraphics[width=0.49\textwidth]{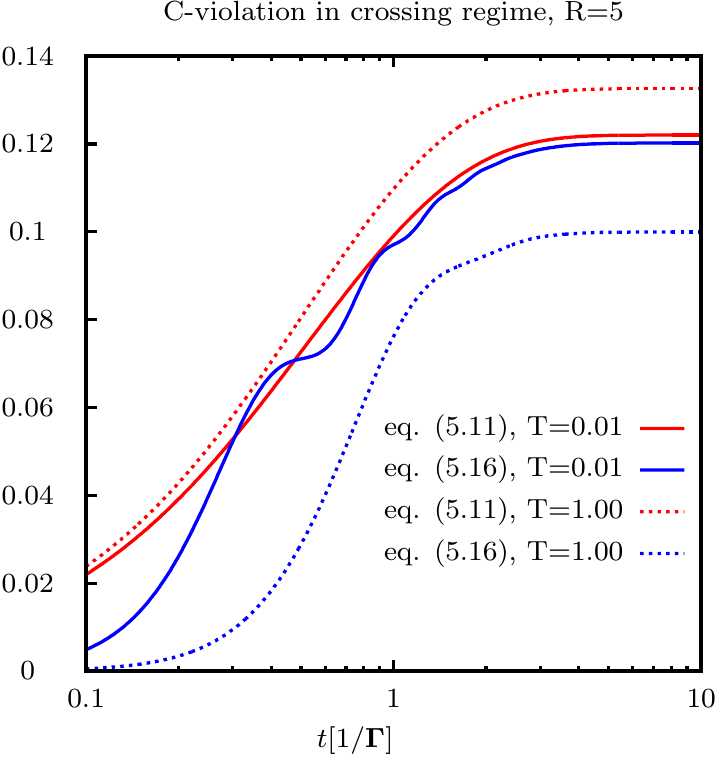}
\caption{\label{tDependence}Time dependence of the two approximate expressions for $\tr\,\eta$ 
(in units of $1/\mu$)
in the runaway and crossing regimes for $R=5$ and two values of the temperature, $T=0.01$ and 
$T=1$ (in units of $\mu$). Time is given in units of $1/\Geff\equiv 1/\Geff_1+1/\Geff_2$.}
\end{figure}

For illustration we present the time-dependence of the 
two expressions for $\tr\,\eta$, \eqn\eqref{TrBhierarchicalacc} and \eqn\eqref{TrBQuasiDegAnalyt} 
in figure \ref{tDependence},
for $R=5$. For this value of the degeneracy parameter both the hierarchical and quasi-degenerate  
approximations are expected to be reasonably good. 
At low temperatures the two approximations give similar results. On the other hand at high 
temperatures they yield values for the final asymmetry which differ by roughly 40\%.  

\paragraph{Asymptotic behaviour}

For $t\rightarrow \infty$ the hierarchical and quasi-degenerate approximations, 
\eqn\eqref{TrBhierarchicalacc} and \eqn\eqref{TrBQuasiDegAnalyt}, simplify to 
\begin{subequations}
\label{TrBAsympt}
\begin{align}
\label{TrBhierarchicalAsympt}
\tr\,\eta(\infty,\vec{q})   = & -\frac{J}{\det M}\,\frac{|Z|^2}{|x_1-x_2|^2} 
\,\sum_{I=1,2} \frac{\Pi^2_\rho(\omega_I,\vec{q})}{(2 \omega_I)^2} 
\frac{1}{\Geff_I}\,,\\
\label{TrBQuasiDegAsympt}
\tr\,\eta(\infty,\vec{q})   =  & -\frac{J}{\det M}\,\frac{|Z|^2}{|x_1-x_2|^2}
\frac{\Pi^2_\rho(\bar{\omega}_{\vec{q}},\vec{q})}{(2 \bar{\omega}_{\vec{q}})^2} 
\nonumber\\
&\times \,\biggl[
\sum_{I=1,2}\frac{1}{\Gamma_{I}}
-2\Re \frac{1}{i(\omega_1-\omega_2)+\frac12(\Geff_1+\Geff_2)}\biggr]\,.
\end{align} 
\end{subequations}
We can try to improve the estimate of the asymptotic value of the asymmetry. To this end 
we note that for $t\rightarrow \infty$ the contributions of the oscillating terms in 
\eqref{TrBHierarchical} and \eqref{TrBQuasiDegAnalyt1} vanish. The integrations over $p_0$ 
and $k_0$ can then be performed analytically:
\begin{subequations}
\label{Exactp0k0Int}
\begin{align}
\label{Exactp0Int}
\int^\infty_0\frac{dp_0}{2\pi}&\frac{1}{p_0^2-q^2_{0,I}} \frac{1-e^{i(q_0-p_0)t}}{q_0-p_0}
\rightarrow \lim_{\epsilon\rightarrow 0}\int^\infty_0 \frac{dp_0}{2\pi} 
\frac1{(p^2_0-q^2_{0,I})} \frac{1}{q_0-p_0+i\epsilon}\nonumber\\
&=\frac1{2\pi}\left(
\frac{\ln(q_0)-i\pi}{(q^2_0-q^2_{0,I})}
+\frac1{2q_{0,I}}\frac{\ln(q_{0,I})}{q_0+q_{0,I}}
-\frac1{2q_{0,I}}\frac{\ln(-q_{0,I})}{q_0-q_{0,I}}\right)\,,\\
\label{Exactk0Int}
\int^\infty_0\frac{dk_0}{2\pi}&\frac{1}{(k_0^2-q^2_{0,I})^*}\frac{1-e^{-i(q_0-k_0)t}}{q_0-k_0}\rightarrow\lim_{\epsilon\rightarrow 0}\int^\infty_0  \frac{dk_0}{2\pi} 
\frac1{(k^2_0-q^2_{0,I})^*} 
\frac{1}{q_0-k_0-i\epsilon}\nonumber\\
&=\frac1{2\pi}\left(
\frac{\ln(q_0)+i\pi}{(q^2_0-q^2_{0,I})^*}
+\frac1{2q^*_{0,I}}\frac{\ln(q^*_{0,I})}{q_0+q^*_{0,I}}
-\frac1{2q^*_{0,I}}\frac{\ln(-q^*_{0,I})}{q_0-q^*_{0,I}}\right)\,.
\end{align}
\end{subequations}
A numerical comparison of the $q_0$ integrands computed using \eqref{Approxp0k0Int} and 
\eqref{Exactp0k0Int} shows that in the relevant range of $q_0$ the difference between
the two is completely negligible even at the crossing 
point. In other words, \eqref{TrBhierarchicalAsympt} provides a very accurate estimate 
of the asymmetry for a strongly hierarchical mass spectrum, whereas \eqref{TrBQuasiDegAsympt}
provides a very accurate estimate of the asymmetry for a quasi-degenerate mass spectrum. 
Their temperature dependence in the runaway and crossing regimes for $R=1$ is presented in figure 
\ref{TrBRdependence}. 
\begin{figure}
    \includegraphics[width=0.49\textwidth]{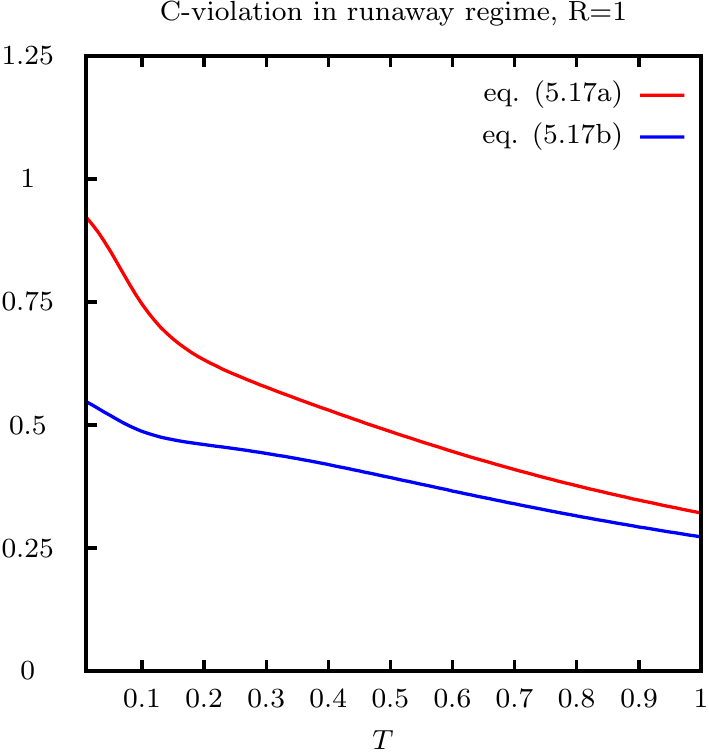}
    \,\,
    \includegraphics[width=0.49\textwidth]{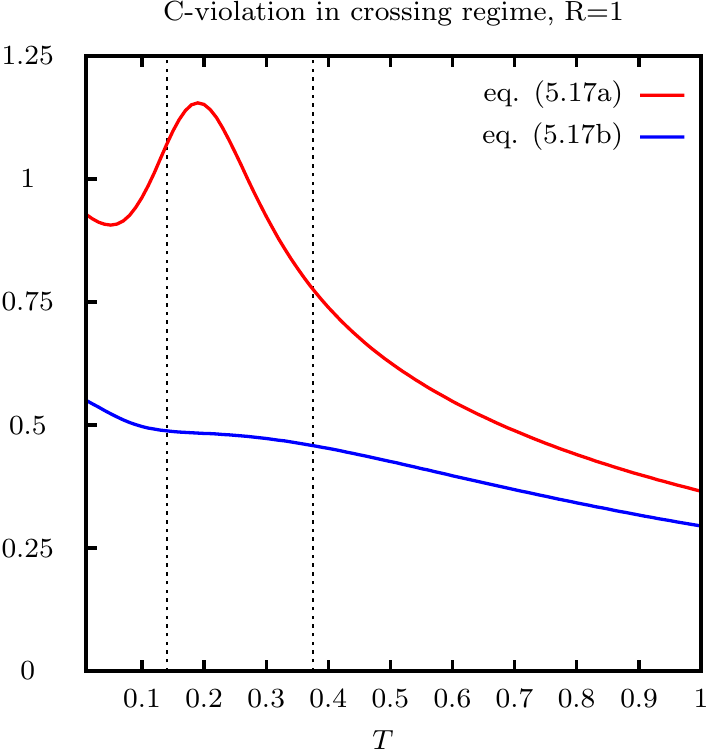}
\caption{\label{TrBTdependence} Dependence of the asymptotic value of $\tr\,\eta$  (in units of 
$1/\mu$) on the temperature (in units of $\mu$) in the runaway and crossing regimes for $R=1$.}
\end{figure} 
In the crossing regime the hierarchical approximation for $\tr\,\eta$ develops a spurious peak. 
Contrary to what one would naively expect the peak is located at a temperature somewhat lower 
than the crossing temperature. This can be traced back to the second term in the denominator
of \eqref{EpsilonMathias}. The effective widths also depend on the temperature, see figure 
\ref{EffMassesWidths}. For $R=1$ the difference of the effective widths reaches minimum around 
$T\sim 0.14$, whereas the difference of the effective masses reaches minimum around $T\sim 0.38$.
As a result $\tr\,\eta$ computed in the Boltzmann approximation peaks at a temperature between 
these two. The contribution of the negative interference terms ensures that the quasi-degenerate 
approximation for $\tr\,\eta$ remains smooth for all values of $T$ and does not develop a peak.

For intermediate values of $R$ neither   \eqref{TrBhierarchicalAsympt} nor 
\eqref{TrBQuasiDegAsympt} is applicable. To obtain an analytical expression valid for any $R$ we note that 
in the $p_0$ integrand in \eqref{TrBEffMGammaAppr} it is sufficient to approximate $\Pi_R(p_0,\vec{q})$ 
by $\Pi_R(\omega_I,\vec{q})$. Furthermore, in the relevant $q_0$ range $\Pi_\rho(q_0,\vec{q})$ 
is a smooth function of $q_0$ and is well approximated by its linear expansion. The integration 
then results in 
\begin{align}
\label{TrBAsymptImproved}
\tr\,\eta(\infty,\vec{q})   = & -\frac{J}{\det M}\,\frac{|Z|^2}{|x_1-x_2|^2} 
\,\biggl[\sum_{I=1,2} \frac{\Pi^2_\rho(\omega_I,\vec{q})}{(2 \omega_I)^2} 
\frac{1}{\Geff_I}\nonumber\\
&-\frac12\frac{\Pi_\rho(\omega_1,\vec{q})+\Pi_\rho(\omega_2,\vec{q})}{(2\omega_1)(2\omega_2)}
\,\Re\frac{\Pi_\rho(\omega_1,\vec{q})(1-i\delta)+\Pi_\rho(\omega_2,\vec{q})(1+i\delta)}{i(\omega_1-\omega_2)+\frac12(\Geff_1+\Geff_2)}\nonumber\\
&-\frac14\frac{\Pi_h(\omega_1,\vec{q})-\Pi_h(\omega_2,\vec{q})}{(2\omega_1)(2\omega_2)}
\,\Im\frac{\Pi_\rho(\omega_1,\vec{q})(1-i\delta)+\Pi_\rho(\omega_2,\vec{q})(1+i\delta)}{i(\omega_1-\omega_2)+\frac12(\Geff_1+\Geff_2)}\biggr]\,,
\end{align} 
where $\delta\equiv \frac12(\Geff_1-\Geff_2)/(\omega_1-\omega_2)$. For a strongly hierarchical
spectrum the last two terms are negligible and \eqref{TrBAsymptImproved} reverts to 
\eqref{TrBhierarchicalAsympt}. On the other hand, for a quasi-degenerate mass spectrum we 
can neglect the difference between $\omega_2$ and $\omega_1$ and replace them by $\bar\omega$
in $\Pi_\rho$ and $\Pi_h$. The $\delta$-terms then cancel out, the last line vanishes and \eqref{TrBAsymptImproved} reverts 
to \eqref{TrBQuasiDegAsympt}. The $R$ dependence of $\tr\,\eta$ for various values of the 
temperature is presented in figure 
\ref{TrBRdependence}. 
\begin{figure}
	\animategraphics[width=0.5\textwidth,poster=last]{4}{figures/TrBRdepRA/TrBRdepRAZ}{1}{100}
	\animategraphics[width=0.5\textwidth,poster=last]{4}{figures/TrBRdepCR/TrBRdepCRZ}{1}{100}
\caption{\label{TrBRdependence} Dependence of the asymptotic value of $\tr\,\eta$  (in units of 
$1/\mu$) on the degeneracy parameter $R$ in the runaway and crossing regimes for various values 
of the temperature (in units of $\mu$). In the PDF-version click on the 
figure to play the animation (supported by Acrobat Reader X).}
\end{figure} 
In both regimes $\tr\,\eta$ vanishes at very large and very small values of the degeneracy 
parameter because for large $R$ the \C-violating parameter is proportional to $1/R$ whereas
for small $R$ it is proportional to $R$. Let us stress once again that $\tr\,\eta$ automatically
vanishes in the limit $M_2=M_1$, i.e.~for $R=0$, because the Lagrangian is \C-symmetric in 
this case. Just as expected, \eqn\eqref{TrBAsymptImproved} gives an accurate result for the final 
asymmetry not only for a hierarchical or quasi-degenerate spectrum, but also for the intermediate 
values of $R$. On the other hand, the expression obtained assuming a hierarchical mass spectrum, 
\eqn\eqref{TrBhierarchicalAsympt}, overestimates the final asymmetry at small $R$, whereas the 
expression obtained assuming a quasi-degenerate mass spectrum overestimates the final asymmetry at 
large $R$. In the runaway regime both the hierarchical and quasi-degenerate approximations are
smooth for all values of $R$ and give similar results. On the other hand, in the crossing 
regime the hierarchical approximation for $\tr\,\eta$ develops a spurious peak that we have 
already observed at figure \ref{TrBTdependence}. The contribution of the negative interference 
terms ensures that the exact result for $\tr\,\eta$ remains smooth for all 
values of $R$ and $T$ and does not develop a peak in the vicinity of the crossing point.

\section{\label{Numerics}Numerical treatment of sub-leading effects} 

In this section we compare the contribution of off-shell effects in the late time limit by 
computing numerically the contribution of the second, third and fourth terms in \eqref{TrB4}. 
Thereby we show that the assumptions made in the previous  section are justified. To this 
end we rewrite \eqref{TrB4}, using \eqref{Fdef}, as
\begin{align}
\label{TrB4Num1}
\tr\,\eta(t,\vec{q})= -\frac{2 J}{\det M}\,
& \int^\infty_0\frac{\Pi_\rho(q_0,\vec{q})}{(2\pi)^3}\, \Im\bigl(I_1 I_2 -I_3 I_4 + I_1 I_5 -I_3 I_6\bigr)\,dq_0\,,
\end{align} 
with 
\begin{align}
\label{TrB4NumIntegrals}
I_1 & = \int^\infty_0{dp_0}\frac{(1-e^{i(q_0-p_0)t})\Pi_R(p_0,\vec{q})}{(q_0-p_0)\det\Omega_R(p_0,\vec{q})}\,,\quad\hspace{2.5mm} I_2 = \int^\infty_0{dk_0}\,\frac{1-e^{-i(q_0-k_0)t}}{ (q_0-k_0) \det \Omega_A(k_0,\vec{q})}\,, \nonumber\\
I_3 & = \int^\infty_0{dp_0}\frac{(1-e^{-i(q_0+p_0)t})\Pi_R(p_0,\vec{q})}{(q_0+p_0)\det\Omega_R(p_0,\vec{q})}\,,\quad I_4= \int^\infty_0{dk_0}\,\frac{1-e^{i(q_0+k_0)t}}{ (q_0+k_0) \det \Omega_A(k_0,\vec{q})}\,,\nonumber\\
I_5 & = \int^\infty_0{dp_0}\frac{(1-e^{-i(q_0-p_0)t})}{(q_0+p_0)\det\Omega_R(p_0,\vec{q})}\,,\quad \hspace{5mm} I_6= \int^\infty_0{dk_0}\,\frac{1-e^{i(q_0-k_0)t}}{ (q_0-k_0) \det \Omega_A(k_0,\vec{q})}\,,
\end{align}
The different integrals $I_1 \ldots I_6$ are all complex valued and finite. In particular 
the integrands are finite at $q_0 = p_0\,,k_0$. The integrands of $I_1\,,I_3\,,I_5$ exhibit 
integrable singularities for $p_0 = |\vec{q}|$ due to logarithmic singularities of the 
self-energies. As discussed below equation \eqref{TrB4}, the contribution of the first 
term dominates the total of \eqref{TrB4Num1} and setting the last three terms in $I_3 \ldots I_6$ 
to zero would correspond to the approximation used in the previous section. Here we are 
interested in their relative contribution in order to know how accurate our assumptions were.  

At finite time the contributions to the integrands by the exponentials are oscillating as 
functions of the integration variables (with frequency  $t$). Since 
the integrals extend to $+\infty$, a large simplification is achieved if we restrict ourselves 
to the late-time limit. Similar to \eqref{Exactp0k0Int}, it is obtained by substituting 
$q_0\rightarrow q_0\pm i\epsilon$ in each of the integrals. We obtain for $t\rightarrow\infty$:
\begin{align}
\label{TrB4NumIntegrals1}
& I_1 = \lim_{\epsilon\rightarrow 0}\int^\infty_0\frac{\Pi_R(p_0,\vec{q})\,{dp_0}}{(q_0-p_0+i\epsilon)\det\Omega_R(p_0,\vec{q})}\,,\quad 
I_2 = \lim_{\epsilon\rightarrow 0}\int^\infty_0\,\frac{{dk_0}}{ (q_0-k_0-i\epsilon) \det \Omega_A(k_0,\vec{q})}\,, \nonumber\\
& I_3 = \int^\infty_0\frac{\Pi_R(p_0,\vec{q})\,{dp_0}}{(q_0+p_0)\det\Omega_R(p_0,\vec{q})}\,,\quad 
I_4= \int^\infty_0\,\frac{{dk_0}}{ (q_0+k_0) \det \Omega_A(k_0,\vec{q})}\,,\nonumber\\
& I_5 = \int^\infty_0\frac{{dp_0}}{(q_0+p_0)\det\Omega_R(p_0,\vec{q})}\,,\quad 
I_6= \lim_{\epsilon\rightarrow 0}\int^\infty_0\,\frac{{dk_0}}{ (q_0-k_0+i\epsilon) \det \Omega_A(k_0,\vec{q})}\,.
\end{align}
Technically we choose a small constant value for $\epsilon$ in the numerical computation which we 
decrease until the results do not change significantly anymore. The performed transformations have 
turned \eqref{TrB4Num1} into a two-fold integral. However the hermitian self-energy included in 
$\Pi_R$ contains a further integral such that the overall dimensionality of $\tr\,\eta(t,\vec{q})$ 
is 3. The quadrature is complicated by the presence of the integrable singularity of $\Pi_R$ at 
$p_0=\left|\vec{q}\right|$ and the poles due to $\det\Omega_R$ and $\det\Omega_A$. These difficulties 
can be overcome by performing appropriate integral transformations on small intervals surrounding these poles.

\begin{figure}
\begin{center}
    \includegraphics[width=0.49\textwidth]{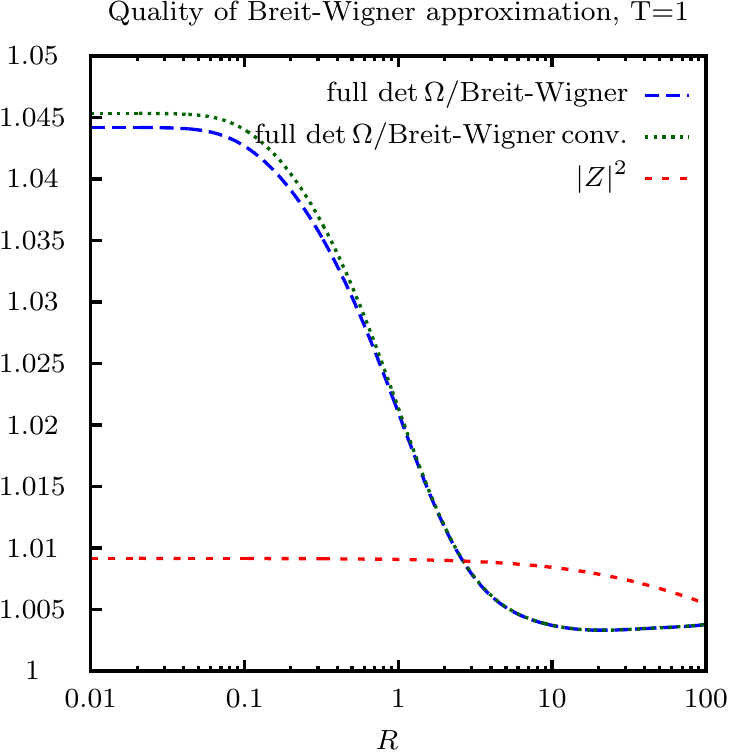}
\end{center}
\caption{\label{TrBRBreitWignerQuality}  Dependence of the ratio of $\tr\,\eta(\infty,\vec{q})$, computed 
without Breit-Wigner approximation, on the degeneracy parameter in the crossing regime for 
$T=1$ (in units of $\mu$). Also shown is the value of $\left|Z\right|$ which 
enters the result in the Breit-Wigner approximation. The $R$-dependence in the 
runaway regime is similar and not shown. `Breit-Wigner conv.' refers to 
the conventional method to determine effective masses and widths sketched above in which one searches 
the complex zeros of $\det\Omega$.}
\end{figure}

Further simplification is achieved by performing the Breit-Wigner approximation in
\eqref{TrB4NumIntegrals1}, i.e.~by using \eqref{detOmegaRdec} and \eqref{detOmegaAdec} everywhere. In 
this approximation it is possible to obtain closed-form analytic expressions for $I_1\ldots I_6$ 
neglecting the finite temperature contributions to $\Pi_R$. At non-zero temperature these 
contributions to $I_1$ and $I_3$ still need to be integrated numerically but the partial analytic 
results stabilize the computation of the total integral \eqref{TrB4Num1}. To demonstrate the accuracy 
of the Breit-Wigner approximation we present the ratio of the first term of \eqref{TrB4Num1}, 
computed without and with this assumption, in figure \ref{TrBRBreitWignerQuality}. We find that 
the relative error made for average momenta is below $5$\%, even for 
smallest degeneracy parameters.

\begin{figure}
    \includegraphics[width=0.49\textwidth]{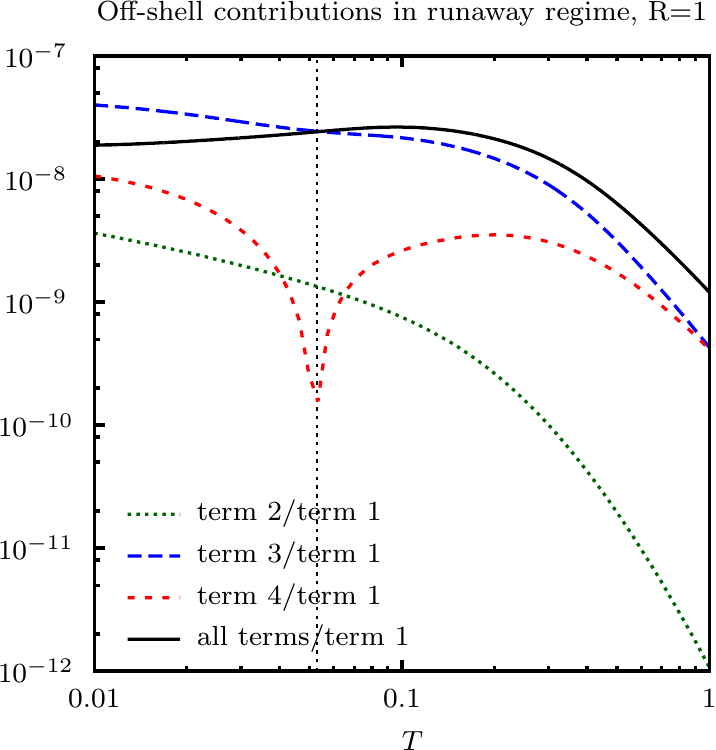}
    \,\,
    \includegraphics[width=0.49\textwidth]{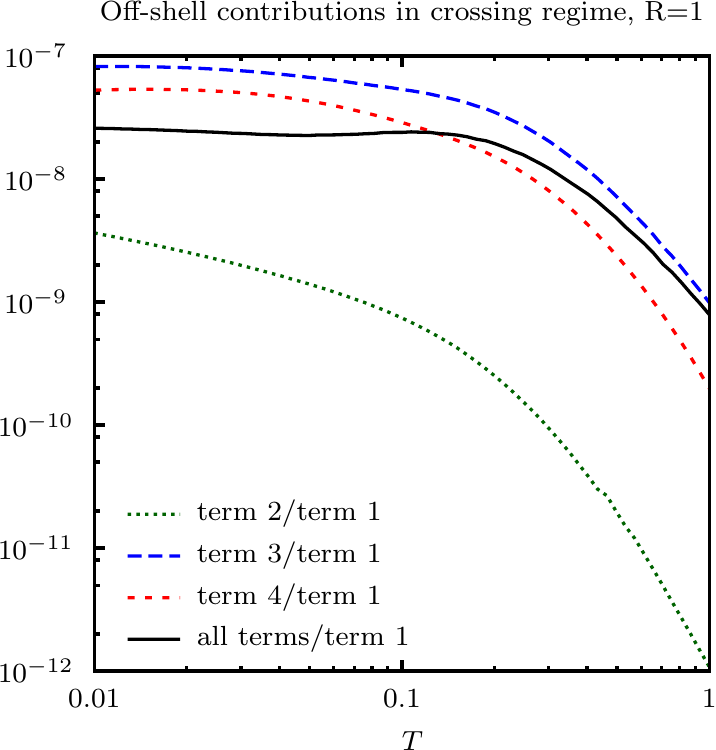}
\caption{\label{TrBTdependenceOffshell} Ratio of the asymptotic values of the contributions of the 
second, third and fourth term  of \eqref{TrB4} to that of the the first term. Shown is the dependence 
on the temperature (in units of $\mu$) in the runaway and crossing regimes for $R=1$. The contributions 
from term 3 and term 4 have opposite sign for $T$ below the dotted vertical line.}
\end{figure}

This result motivates us to study the contribution of off-shell effects in the Breit-Wigner 
approximation as well. Figure \ref{TrBTdependenceOffshell} shows the $T$-dependence of the relative 
contributions by the second, third and fourth term to \eqref{TrB4Num1} with respect to that by the 
first one. The contributions of the third and fourth terms are small, for all values of $T$, compared 
to the on-shell contribution because only one of the $(q_0\pm k_0)$ and $(q_0\pm p_0)$ factors in 
the denominators of the integrands of $I_1,I_3,I_5,I_6$ can vanish for each of them.  For the same 
reason the contribution of the second term is even smaller. In this case none of the $(q_0+ k_0)$ 
and $(q_0+ p_0)$ terms can vanish since the integration is only over positive momenta. Additionally 
the two contributions by the third and the fourth term  enter with opposite signs (only for small 
$T$ in case of the runaway regime).

\begin{figure}
    \includegraphics[width=0.49\textwidth]{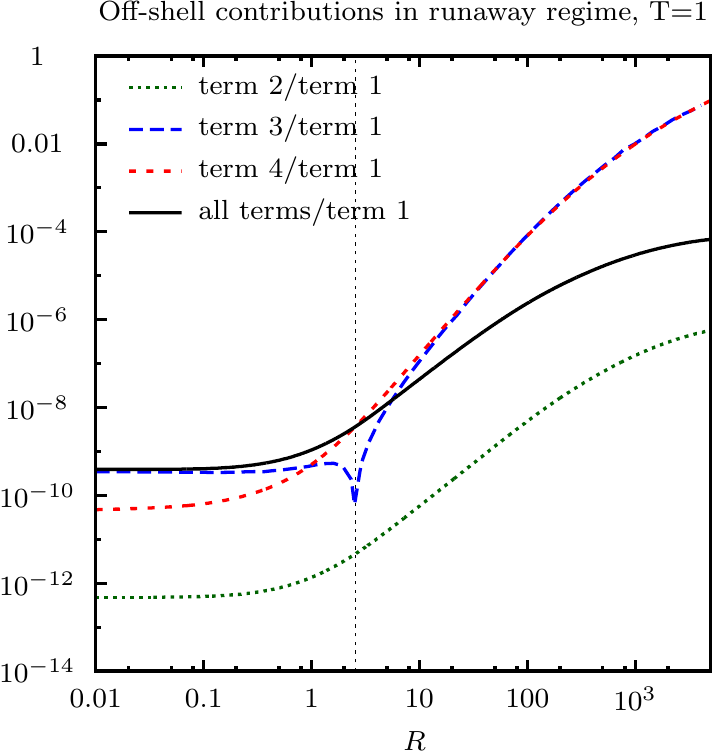}
    \,\,
    \includegraphics[width=0.49\textwidth]{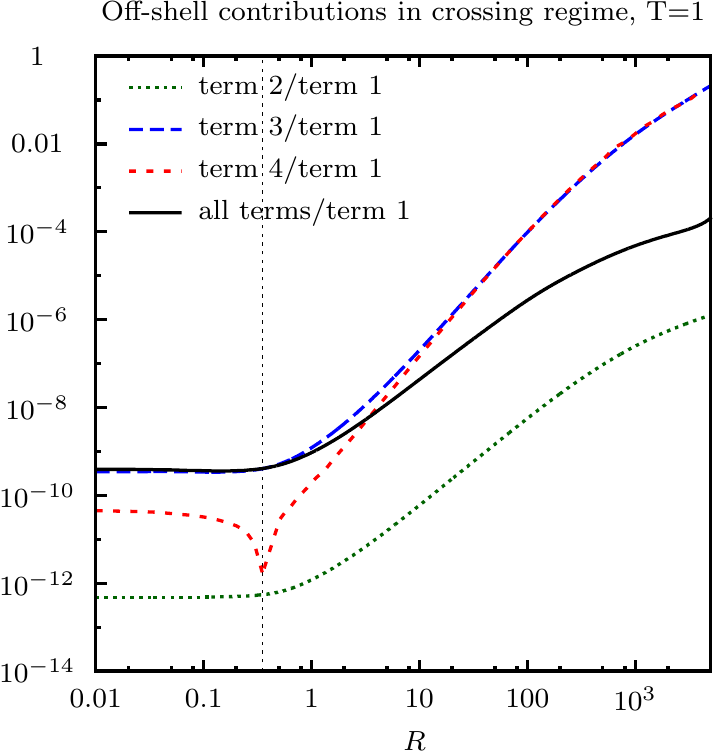}
\caption{\label{TrBRdependenceOffshell} Ratio of the asymptotic values of the contributions of the 
second, third and fourth term of \eqref{TrB4} and that of the the first term. Shown is the dependence 
on the degeneracy parameter in the runaway and crossing regimes for $T/\mu=1$. The contributions 
from term 3 and term 4 have opposite sign for $R$ above the dotted vertical line.}
\end{figure}

Figure \ref{TrBRdependenceOffshell} shows the $R$-dependence of the same quantities for fixed 
temperature $T=1$ (in units of $\mu$) and the corresponding average momentum $\vec{q}$. Remarkably 
the relative contribution of off-shell effects increases with increasing degeneracy parameter. It 
flattens for large $R$ and stays below $1$\% in our examples. For this behaviour it is crucial that 
the contributions of the third and fourth terms of \eqref{TrB4Num1} cancel again for large $R$.

\section{\label{Summary}Summary and outlook}

For resonant leptogenesis the mass difference is typically of the 
order of the sum of the decay widths. Therefore the thermal corrections to the effective masses, 
which are of the order of the widths, are 
comparable to the mass difference. Depending on the values of the couplings either runway or 
crossing regime is realized. These have not been discussed in the context of leptogenesis before. 
In the runaway regime the mass difference grows with increasing 
temperature, whereas in the crossing regime the difference of the masses initially decreases, 
such that the effective masses become equal at some temperature, and then increase again at even 
higher $T$. The main goal of this work was to investigate the asymmetry generation 
in the vicinity of the crossing point. We obtained consistent results for the \CP-violating 
source term which maintains the \CP-properties of the Lagrangian as the temperature changes. 
In particular this enabled us to find answers to three questions which may be asked based on 
the results of previous studies  relying on Boltzmann-like equations combined with conventional 
quantum field theory: Is the source term suppressed (or does it even change sign) when the 
difference of the effective masses vanishes at the crossing point? Can a Boltzmann-like 
approximation, which takes into account thermal corrections to the masses and decay widths, 
adequately describe the asymmetry generation close to a crossing point? What 
is the relative size of off-shell contributions beyond the Breit-Wigner approximation?

As far as the first question is concerned,
it has been found in earlier works that the \CP-violating parameters are proportional to 
the mass difference. In the limit of equal masses they vanish (as the mass difference passes 
through zero)  which is consistent with the  
\CP-invariance of the Lagrangian in this case. Because the difference of the effective masses 
vanishes at the crossing temperature, one could naively expect that \CP-violation also vanishes 
or is at least suppressed at the crossing point. Here we have demonstrated analytically that 
the masses in the numerator and denominator of \eqref{EpsilonVacuum} have different origin.
Whereas masses and widths in the denominator of the canonical expression for the \CP-violating 
parameter may be interpreted as the
effective thermal masses, the numerator contains the mass parameters of the Lagrangian. These do 
not depend on temperature by definition. Therefore, contrary to the naive expectation,
the vanishing of the difference of the effective 
masses by no means implies vanishing of the \CP-violating source term. The ability to distinguish 
between the 
mass parameters of the Lagrangian and the effective masses relies on an important 
technical aspect of our analysis. In contrast to the approach followed in earlier 
works we did not use  renormalization group equations to minimize the difference 
between the mass parameters and effective masses at each temperature. Instead we 
have fixed the renormalization scale at zero temperature. As a result,  
the mass matrix appearing in the Kadanoff-Baym equations and throughout the rest
of the paper coincides with the mass parameters of the Lagrangian. The latter can 
in principle be measured experimentally at zero temperature in e.g. decay and scattering 
experiments. In other words, this approach has the advantage 
that the relation between the, in principle measurable, 
zero-temperature masses and widths and the asymmetry generated at temperatures 
comparable to the masses remains transparent. 

Concerning the second question,
 peaks of the spectral functions that correspond to the quasiparticle 
excitations may strongly overlap in the resonant regime. This renders the applicability of the 
Boltzmann approximation questionable. One could  expect that this approximation breaks down 
completely at the crossing point. Our analysis confirms that even taking the thermal effects in the 
form of effective masses and widths into account does not substantially improve the quality
of the quasiparticle approximation. In particular, close to the crossing point the asymmetry 
computed in the Boltzmann approximation develops a spurious peak absent in the exact result. 
Additional contributions that describe coherent transitions between the 
two mass shells exactly compensate the enhancement of the non-oscillating contributions. We 
would like to stress that these coherent transitions are determined by basis-invariant 
effective masses and widths and affect all components of the two-point functions, which are 
not basis-invariant. In other words, the dynamics is \emph{basis-covariant} and can be 
formulated in terms of \emph{basis-invariant} quantities.  Because in both regimes the mass 
difference grows at high temperatures one can expect that the quality of quasiparticle
approximation  increases. We find that this is indeed the case. However, this 
improvement is not related to an increasing separation of the peaks of the spectral function. 
It turns out that for a quasi-degenerate mass spectrum at high temperatures the positions of the 
peaks are not determined by the effective masses. 

Finally, coming to the third question,
the very fact that in the crossing regime the quasiparticle approximation completely breaks down
also raises the question of the relative size of off-shell contributions. Comparing our 
analytical results with exact numerical computations we find that purely off-shell effects 
are small, of the order of $1$\% or less for our benchmark scenarios. The Breit-Wigner approximation 
itself entails a relative error of less than $5$\% even at the crossing point. Note that all 
our computations relied on the toy-model and that quantitative results could differ for the 
phenomenological scenario.

In addition to clarifying the qualitatively important and interesting questions raised above, 
the results of this work also provide a reference solution for various approximation schemes.
Here we have demonstrated that the Boltzmann approximation is not applicable in the crossing 
regime. In a forthcoming publication we will study the applicability of the density matrix 
equations in the two regimes and present a derivation of the density matrix equations from 
first-principles.

\section*{Acknowledgements}

We would like to thank M. Shaposhnikov for valuable discussions. This research was  
supported in part by the National Science Foundation under Grant No. PHY11-25915.

\providecommand{\href}[2]{#2}\begingroup\raggedright\endgroup

\end{document}